\documentclass[preprint,12pt,3p]{elsarticle}

\biboptions{sort&compress}
\usepackage{multicol}
\usepackage{color}
\usepackage{xspace}
\usepackage{enumerate}
\usepackage{lineno}
\usepackage{graphicx}
\graphicspath{{./figures/}}
\usepackage{float}
\usepackage{setspace}
\usepackage{subcaption}
\usepackage{wasysym}
\usepackage{amssymb}
\usepackage{multirow}
\usepackage{adjustbox}
\usepackage{siunitx}

\let\today\relax
\makeatletter
\def\ps@pprintTitle{%
    \let\@oddhead\@empty
    \let\@evenhead\@empty
    \def\@oddfoot{\footnotesize\itshape
         {https://doi.org/10.1016/j.nima.2020.164106} \hfill\today}%
    \let\@evenfoot\@oddfoot
    }
\makeatother

\begin{document}
\begin{frontmatter}

\title{Slow Fluors for Effective Separation of  Cherenkov Light \\ in Liquid Scintillators}

\author{Steven D. Biller \corref{cor1}}
\cortext[cor1]{corresponding author}
\ead{steven.biller@physics.ox.ac.uk}
\author{Edward J. Leming}
\author{Josephine L. Paton}
\address{Department of Physics, University of Oxford, Oxford OX1 3RH, UK \\ \vskip 0.2in [NIM A 972 (2020) 164106]
}

\begin{abstract}
The timing and spectral characteristics of four highly efficient, slow fluors are presented for liquid scintillator solutions using linear alkylbenzene (LAB) as the primary solvent. The mixtures exhibit high light yields, but with rise times of several ns or more and decay times on the order of tens of ns. Consequently, such liquid scintillator mixtures can be used foreffective separation of Cherenkov and scintillation components based on timing in large scale liquid scintillation detectors. Such a separation, showing high light yield and directional information, is demonstrated here on a bench-top scale for electrons with energies extending below 1 MeV. This could have significant consequences for the future development of such detectors for measurements of solar neutrinos and neutrinoless double beta decay (0$\nu\beta\beta$) as well as providing good directional information for elastic scattering events from supernovae neutrinos and reactor anti-neutrinos, amongst others.
\end{abstract}


\end{frontmatter}

\section{Introduction}
The potential advantages of detecting the distinct Cherenkov and scintillation components of light signals in large liquid scintillation detectors via time separation has been highlighted by a number of authors (see for example \cite{biller,aberle,li,Guo}). Combining advantages of both detection techniques, the goal would be to use the Cherenkov signal to provide directional and topological information while maintaining the good energy resolution of bright liquid scintillators. Furthermore, the ratio of Cherenkov to scintillation signals could be used to provide information for particle identification and background discrimination. Typical approaches have so far either relied on timing improvements to photodetectors \cite{aberle,hamamatsu} and/or weak scintillators \cite{Yeh}, often using a low concentration of primary fluor to decrease non-radiative transfer and reduce contamination of the Cherenkov signal \cite{Guo,li,LSND}. The difficulty with the former approach is that wavelength dispersion in large detectors will broaden the prompt Cherenkov signal over several ns, limiting the extent of signal separation in typical scintillation mixtures even for ideal photodetectors. Simulation studies have so far only indicated modest signal separation for large instruments \cite{aberle}. The difficulty with the latter approach is that it sacrifices light yield and, hence, energy resolution. A different method recently proposed \cite{dicroic} uses dichroic films to spectrally separate the portion of Cherenkov light above 450nm from the scintillation signal. However, this approach would require potentially costly hardware upgrades and has a Cherenkov photon collection efficiency that is limited by the spectral range and photocathode coverage dedicated to these photons. 

The work presented in this paper instead follows the suggestion of \cite{biller} to develop slow liquid scintillators with high light yields. This would offer a cost-effective approach that could be applied to existing detectors to provide excellent Cherenkov separation over the whole range of photocathode coverage. Directional information could then be used effectively down to much lower energies, which is relevant for topics such as low energy solar neutrinos as well as neutrinoless double-beta decay. While the slower scintillation signal will degrade vertex resolution to some extent, this can be tuned relative to the Cherenkov separation purity by choosing different fluor mixtures. For energies above a couple MeV, the presence of the prompt Cherenkov component in any case constrains this resolution in large-scale detectors to be somewhere between that for standard liquid scintillator ($\sim$10cm) and pure Cherenkov ($\sim$30cm) instruments.

Linear Alkylbenzene (LAB) will be used as the primary solvent for this study as it has been shown to be easy to handle with high intrinsic light yield, excellent optical properties and is either in use or planned to be used by a number of liquid scintillation experiments, for example \cite{DayaBay,RENO,SNO+,JUNO,Jinping}. Four fluors have been selected in this study, two of which (acenaphthene and pyrene) are suitable as primary fluors in LAB, and two (9,10-diphenylanthracene (DPA) and 1,6-diphenyl-1,3,5-hexatriene (DPH)) as secondary fluors. Like all polyaromatic hydroharbons and polyenes, these fluors tend to be light-sensitive to varying degrees, so exposure to UV should be minimised to avoid degradation of scintillator optical quality. While absorption and emission characteristics of these fluors have been measured before \cite{Berlman}, there can be significant solvent effects on emission spectra. It is also important to measure absorption spectra over a large dynamic range relevant for large scale detectors. Measurements of the relevant properties of these fluors in LAB mixtures will therefore be described in the following sections, along with light yield, timing characteristics and demonstrations of directional Cherenkov light separation for electron energies in the region below 1 MeV.

\section{Experimental methods}

\subsection{Light Yield}
The relative light yield of scintillation mixtures was determined using samples in borosilicate scintillation vials irradiated by a $^{90}$Sr source and viewed by a Hamamatsu H11432-100 (SBA) photomultiplier tube positioned $\sim$1cm away from the vial. Duplicate samples were used to assess systematics due to varying vial glass thickness and sample preparation, which dominate the quoted uncertainties. Spectral endpoints were determined based on a pre-defined drop-off in rate after a fixed counting time for each sample. These endpoints were then compared with those from reference samples of LAB (PETRELAB 500-Q from Petresa Canada Inc.) with a 2g/l concentration of 2,5-Diphenyloxazole (PPO), for which an intrinsic light yield of 11900 photons per MeV has previously been established \cite{Penn} based on simulations that account for the LAB absorption and emission spectra along with the detection geometry and PMT efficiency as a function of wavelength. Similarly, the light yields quoted in this paper make use of the quoted quantum efficiency spectral shape for the H11432-100, the measured extinction of LAB as a function of wavelength (Figure~\ref{fig:LAB_Spectrum}) and the measured spectral shape of the fluorescence emission in different samples (presented below) to deconvolve the intrinsic light yield of samples from this relative measurement. These values for the various fluor combinations described are summarised in Table~\ref{tab:fluors}. 

\subsection{Transmission}
Light transmission measurements were made in cyclohexane using quartz cuvettes with a 1cm path length in a Perkins-Elmer Lambda 9000 transmission spectrometer. The absorbance, $A$, is defined as $\log_{10}(I_0/I)$, where $I_0/I$ is the ratio of the radiant energy incident on the sample at a particular wavelength relative to that transmitted by the sample. The molar extinction coefficient was then calculated from the Beer-Lambert Law as $\varepsilon = A/cl$, where $c$ is the molar concentration of the fluor and $l$ is the path-length used. Units for $\varepsilon$ are expressed in liters/mol/cm. For the fluors considered here, a value of $\varepsilon=0.1$ roughly corresponds to an absorption length of 10m at a fluor concentration of $\sim$1 g/l. As this length scale is of relevance for large detectors, efforts have been made to extend the measured range down to this value of $\varepsilon$ for primary fluors. For secondary fluors, relevant concentrations for detectors are typically $\sim$100 times smaller, so measurements down to values of $\varepsilon$=10 are sufficient. A range of fluor concentrations (typically ranging from several milligrams to several grams per litre) was used to cover the range, with attention paid to the consistency of overlapping spectral features.

\subsection{Fluorescence Spectra}
Fluorescence emission spectra were measured with an Andor 303i spectrometer with a 1024x256 back-illuminated CCD and a 300 lines/mm grating, with samples excited by a 266nm UV laser. Samples in quartz vials were illuminated from above via an optical fibre, with the beam spot focussed by optics several mm into the sample. The return light along the same path was then directed into the spectrometer via a part-silvered mirror. All samples were measured in LAB at concentrations similar to those used for time spectra measurements.

\subsection{Emission Time Spectra}
\begin{figure}[H]
    \begin{subfigure}{1.0\textwidth}
    \centering
    \includegraphics[width=0.7\linewidth]{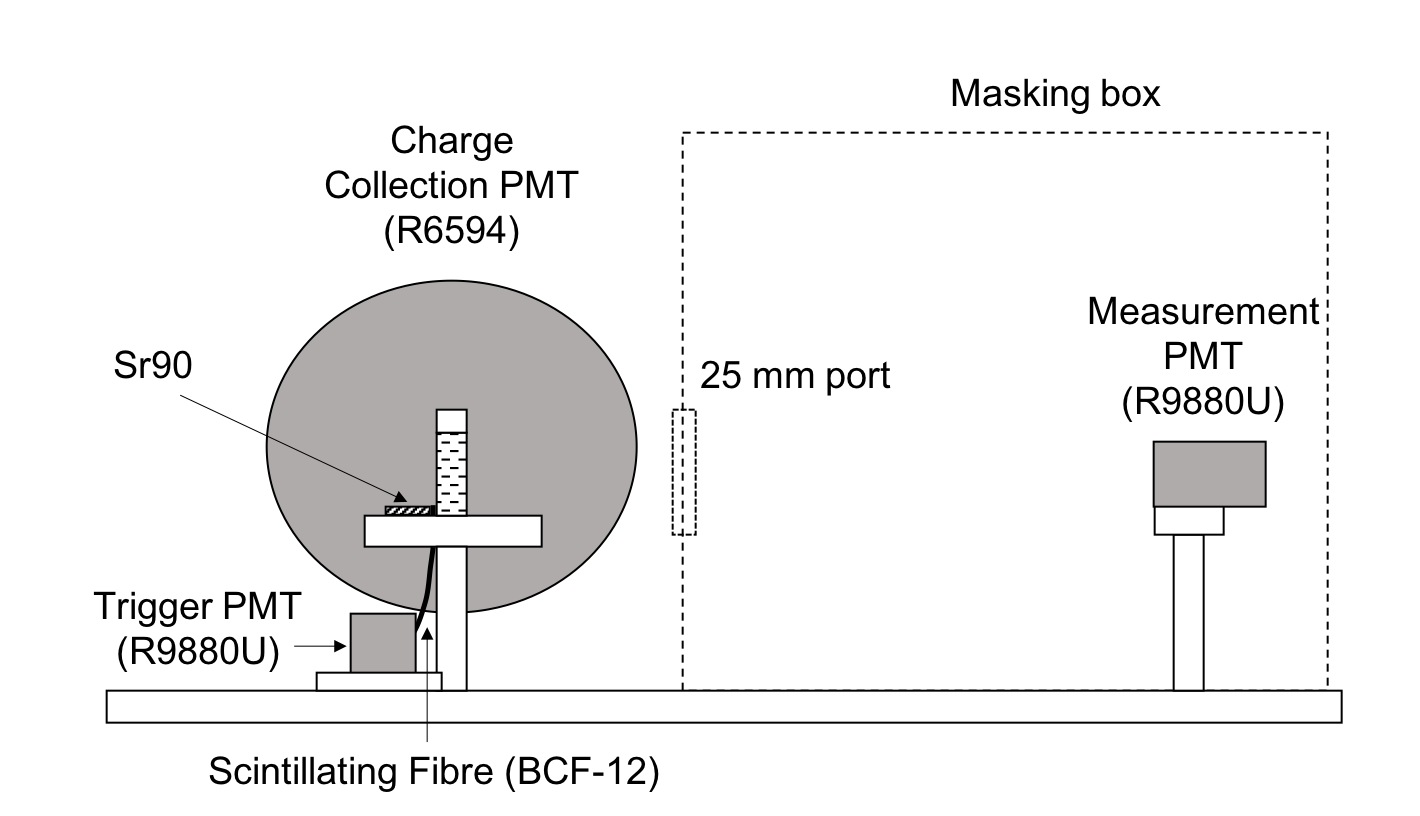}
    \caption{Side view}
    \label{fig:Arrangement_side}
  \end{subfigure}
\newline
  \begin{subfigure}{1.0\textwidth}
    \centering
    \includegraphics[width=0.7\linewidth]{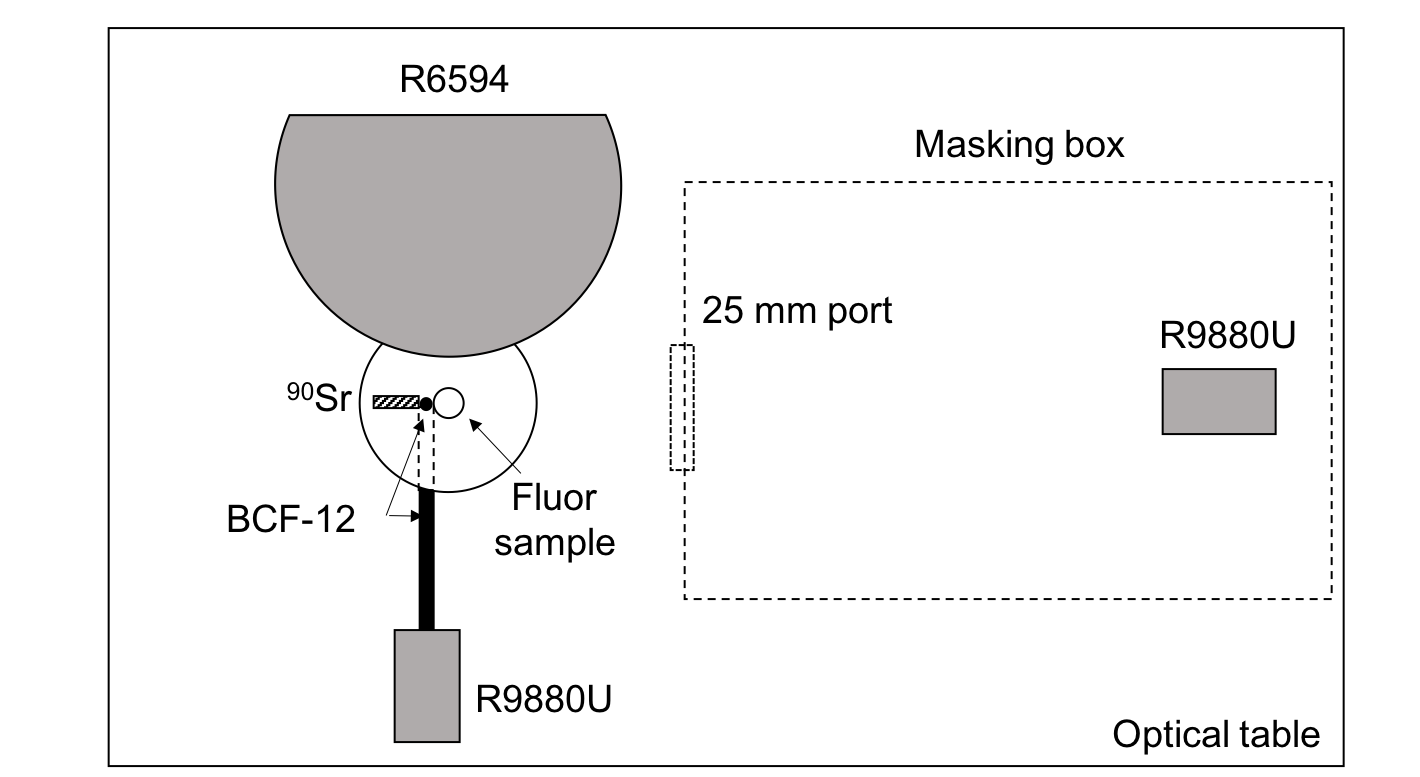}
    \caption{Top view}
    \label{fig:Arrangement_top}
  \end{subfigure}%
  \caption{A diagram showing the experimental arrangement for the \textit{towards} configuration used to measure emission time profiles of slow fluors. The BCF-12 scintillating fibre is fed-through the base plate from below to minimize shadowing}
  \label{fig:ExperimentalArrangement}
\end{figure}

The emission time profiles were determined using an arrangement based on the single photon technique described in \cite{Bollinger} but augmented to provide a trigger that is independent of the fluor under observation. An overview diagram of this arrangement is given in Figure~\ref{fig:ExperimentalArrangement}. The event trigger is provided by a 2~mm diameter Saint-Gobain BCF-12 scintillating fibre optically coupled to a Hamamatsu R9880U-210 (UBA) PMT, henceforth referred to as the \textit{trigger} PMT, using index matching gel. The fibre is fed through the base-plate holding the fluor sample from below and wrapped in highly reflective aluminized foil to maximise light collection. A $^{90}$Sr source, also 2~mm in diameter, is used to excite both the scintillating fibre and the fluor sample inside a Borosilicate glass scintillation vial, which is observed by two PMTs: a Hamamatsu R6594 PMT positioned $\sim$1~cm from the vial and another Hamamatsu R9880U-210 placed $\sim$10-20~cm away. 

The first (\textit{charge collection}) PMT provides a measure of the energy deposited in the scintillator and is used for applying event level selection cuts. The second (\textit{time measurement}) PMT is placed at a distance such that it has less than a 10\% occupancy relative to the charge collection tube. As a result, the number of events in which the measurement tube observes multiple photons is considered negligible and ignored in the analysis. Varying the occupancy by factors of $\sim$2 had no significant impact on the the extracted results, verifing the validity of this approximation. The PMTs, the base-plate that holds the fluor sample and the $^{90}$Sr source were mounted on a Thorlabs optical posts for positional stability. All PMTs were amplified through an ORTEC FTA420 amplifier with a gain of 200 and read out at 2.5 GS/s by a LeCroy Wavepro 7200a oscilloscope. Each digitized waveform is $1~\mu s$ long with 400~ps samples. A coincidence trigger was used to acquire data, enforcing coincident observations between the trigger and measurement PMTs within a 800~ns window.

To minimize the contribution of reflected or scattered photon paths, a masking box was placed over the measurement tube. The box has a 25~mm diameter port on the front face for mounting optical components. When observing pyrene samples, a Thorlabs 450~nm long-pass or an Edmund Optics 400~nm short-pass optical filter was mounted to select the emission components of the excimer and monomer states, respectively. In all other cases, the port housed a Thorlabs ID25 25~mm diameter iris, which was used to fine-tune the occupancy at the measurement tube where required. 

In the results that follow, the orientation of the $^{90}$Sr source, sample and measurement PMT were configured in two arrangements: In the \textit{towards} configuration (Figure~\ref{fig:Arrangement_side}), the source was placed on the far side of the sample so that the Cherenkov light emitted by electrons entering the fluor sample would, on average, be directed \textit{towards} the measurement tube. The \textit{away} configuration oriented the source on the near-side of the sample, such that the Cherenkov light emitted by the electrons was, on average, directed away from the PMT. The whole arrangement was contained within a 120~cm~x~75~cm~x~65~cm sealed dark box. 

To mitigate fluorescence quenching by oxygen, all samples were bubbled with nitrogen before data taking. In order to minimize any ingress of atmospheric oxygen, the vial lid-thread was wrapped in PTFE tape before sealing. 

The $^{90}$Sr source used in these studies supports $^{90}$Y, which undergoes $\beta$-decay with an endpoint of 2.28 MeV. After the emitted electron passes through the 2mm-thick optical fibre and 1mm-thick vial wall, this leaves a maximum energy of $\sim$1.2 MeV that can be deposited in the scintillator sample, though typical energies after event trigger selection are below 1 MeV. For these energies, calculations of the Cherenkov light produced in the non-scintillating vial wall predict a contribution of $\sim$20-30\% to the overall Cherenkov signal observed, depending on the exact electron energy. This was experimentally verified by repeating the acenaphthene measurements of Figure \ref{fig:fit_acenapthene_towards} using an extra glass insert to double the effective vial thickness, comparing cases with and without an opaque layer between the insert and vial to insure the same scintillator pulse height selection. Interestingly, for electrons in this energy range, this fractional glass contribution to the Cherenkov light is comparable to the proportion of their energy deposition expected to fall below the Cherenkov threshold. As a result, the Cherenkov to scintillation ratio indicated by the measurements shown here, which include the glass contribution, is more indicative of what would be seen at higher energies as the fractional deposition below the Cherenkov threshold becomes smaller.

\section{Time profile analysis}
Offline analysis software was used to calculate the time differences between the digitized trigger and measurement signals. All digitized traces were filtered with a 5th order Butterworth infinite impulse response filter with cut-off frequency 500~MHz, 300~MHz and 50~MHz for the measurement, trigger and charge collection signals, respectively. To measure the separation time between trigger and measurement signals, a constant fraction discriminator was implemented in the code to calculate timestamps by linearly interpolating between the sample points that bound a given threshold. A constant fraction of 40\% was used to calculate timestamps at the measurement PMT. For the trigger PMT, timestamps were taken at a 5\% constant fraction threshold, which was found to provide the optimal timing resolution when measuring the impulse response function (IRF) of the system. Charge cuts were applied at both the trigger and charge collection PMTs to remove low energy `tail' events by applying integration windows. The integration windows were centred about leading edge threshold crossings of approximately 2.5 p.e. on the measured transients. For the trigger PMT, the window width was set to [-10, 30] ns and a charge cut of 200~pC (approximately 30 p.e.) was applied.  For the charge collection PMT, the window width was set to [-10, 150] ns and a charge cut of 15~pC (approximately 3 p.e.) was applied.  Any event where multiple crossings of the 2.5 p.e. threshold were observed in either the signal or trigger traces was rejected from the analysis.

\subsection{Measuring the system's impulse response function}

\begin{figure}[H]
\centering
  \begin{subfigure}{0.8\textwidth}
    \centering
    \includegraphics[width=0.9\linewidth]{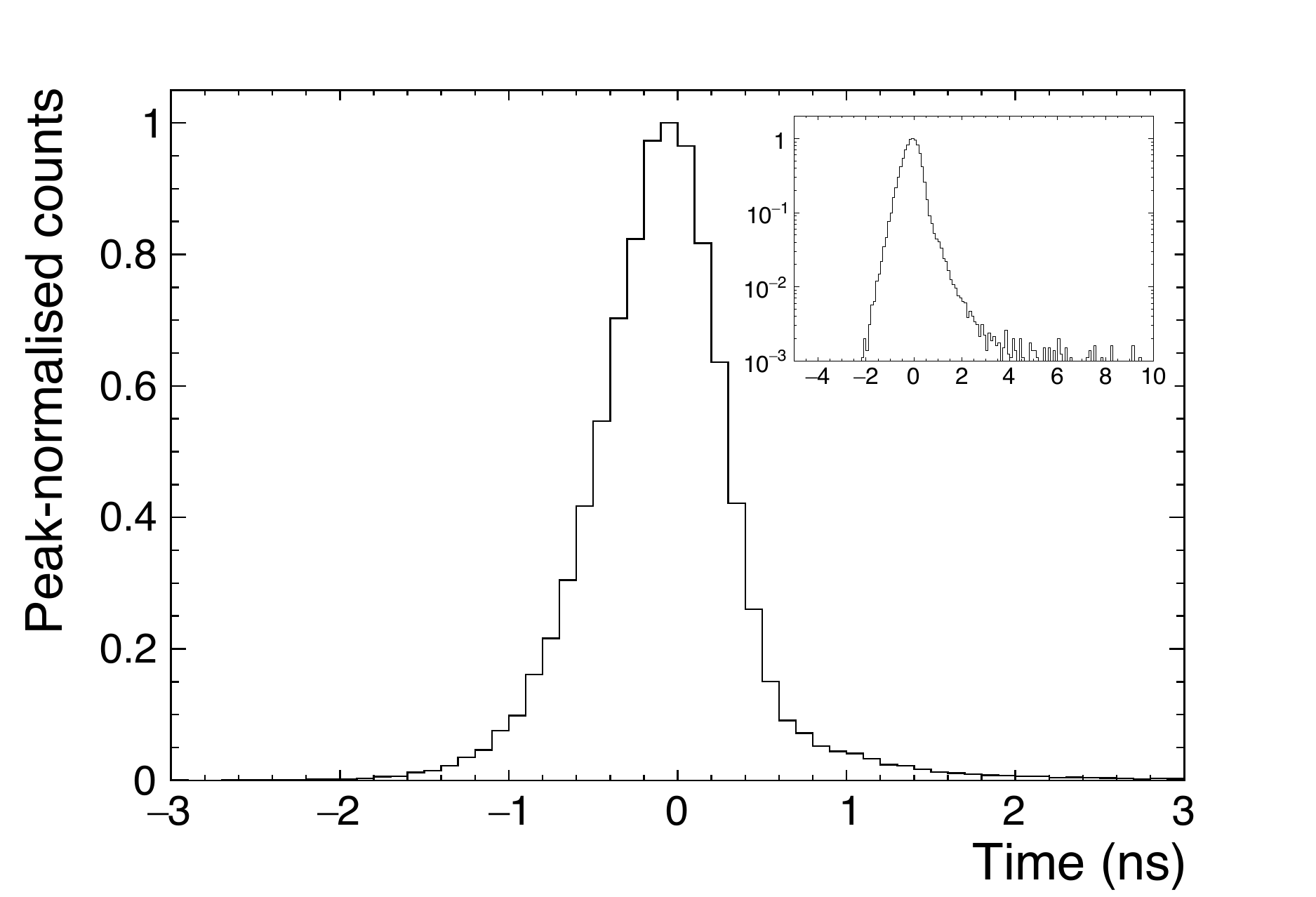}
    \caption{Impulse Response Function}
    \label{fig:IRF}
  \end{subfigure} \\
  \begin{subfigure}{0.8\textwidth}
    \centering
    \includegraphics[width=0.9\linewidth]{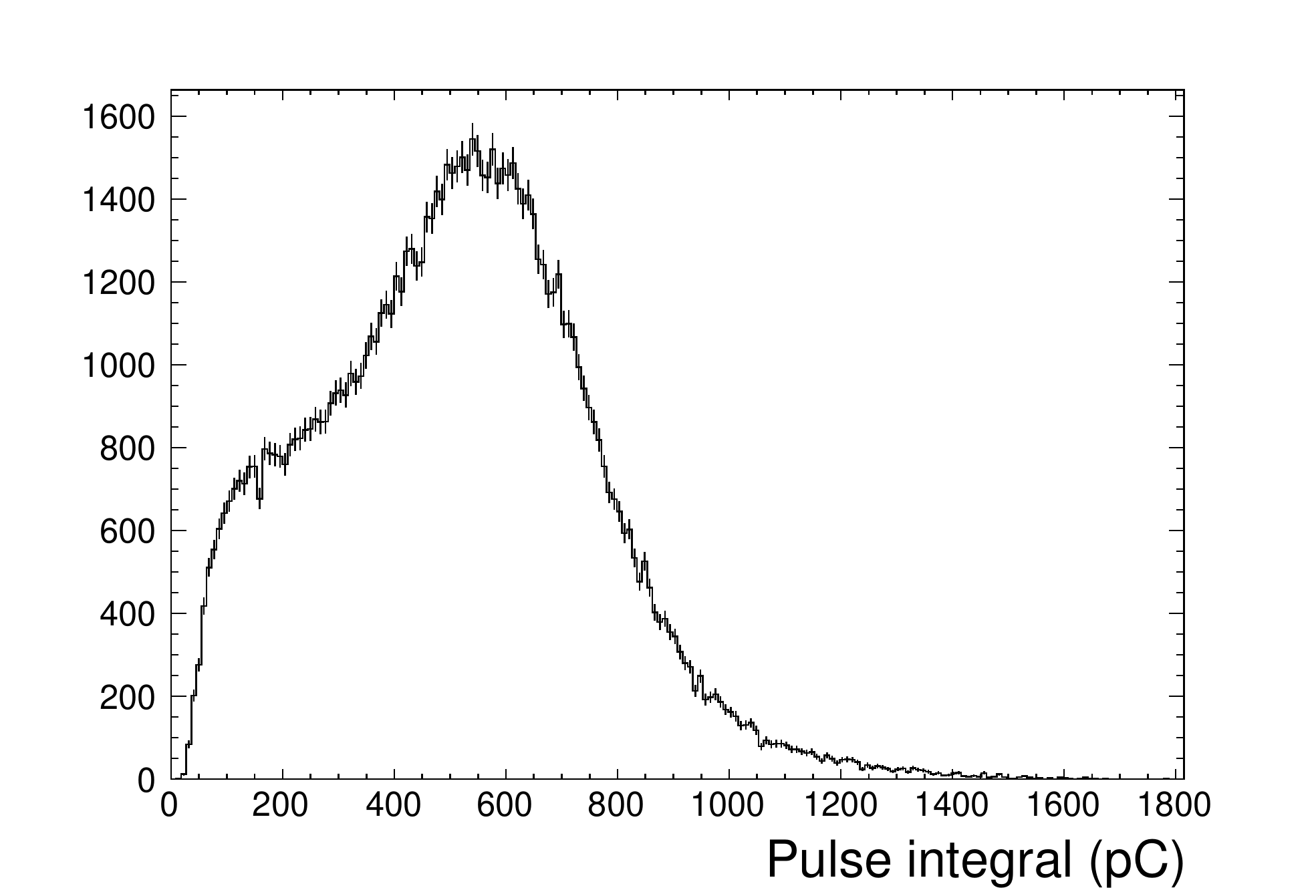}
    \caption{Trigger charge distribution}
    \label{fig:TriggerQ}
  \end{subfigure}
  \caption{The impulse response function of the time-correlated single photon counting set-up using water Cherenkov data: (a) Pulse separation time between trigger and measurement PMTs for a 5\% constant fraction threshold. The inset shows the same distribution on a log scale. The distribution was formed from \SI{1e5} events (b) Charge distribution measured at the trigger PMT.}
\end{figure}

The system IRF was measured by replacing the fluor sample shown in Figure~\ref{fig:Arrangement_top} with a vial of distilled water. In this arrangement, electrons from the $^{90}$Sr source propagate through the trigger fibre and into the water sample, emitting a prompt Cherenkov signal during transit. This prompt signal is modelled as a $\delta$-function response. The IRF of the system is measured by building a histogram of the time difference between the trigger and measurement tubes over a number of events.  As described above, a charge cut of 200~pC was used to reject events in the tail of the trigger PMT's charge distribution. If a Gaussian function is fit to the resulting time distribution given in Figure~\ref{fig:IRF}  (with Figure~\ref{fig:TriggerQ} showing the corresponding charge distribution), the coincident timing resolution ($\sigma$) of the system is found to be approximately 390ps, although the actual measured distribution is explicitly used in the analysis that follows.

\subsection{Fitting procedure}

\begin{figure}[H]
    \centering
    \includegraphics[width=0.8\linewidth]{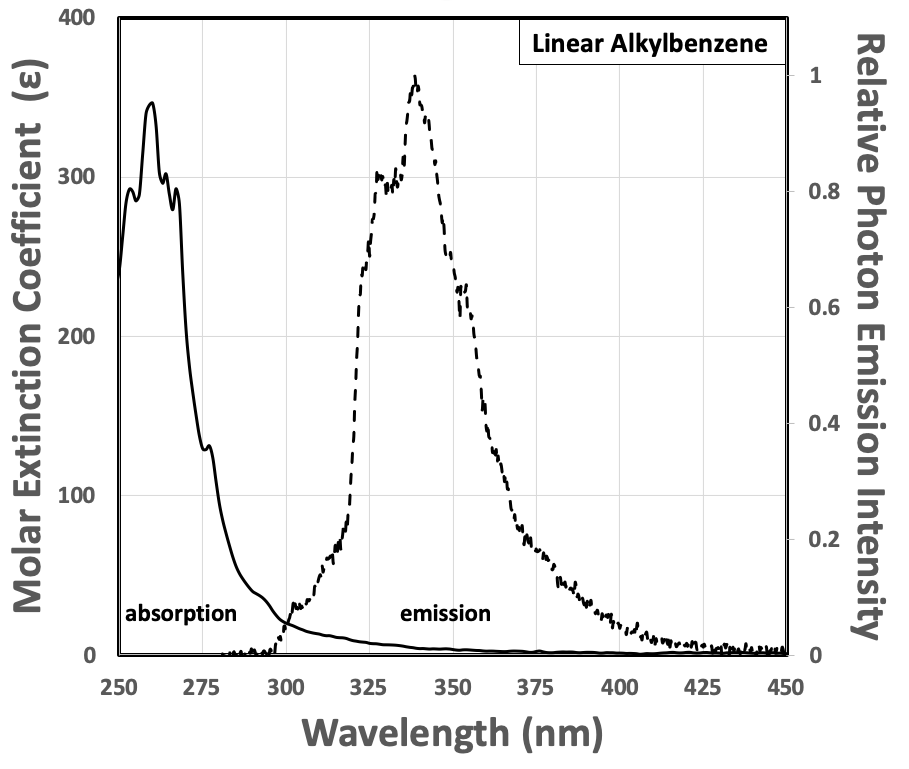}
    \caption{Absorption (solid) and relative emission spectra (dashed) of LAB.}
    \label{fig:LAB_Spectrum}
\end{figure}

The optical response of the slow scintillating fluors was fit using an empirical model consisting of the sum of $n$ exponential decays ($n=1$ or $2$) with a single, common rise time. In all cases aside from pyrene, the fit quality was noticeably improved by including a small additional component with an instantaneous rise time and a fall time ($\tau'$) comparable to the rise time of the added fluor. We believe this can be potentially attributed to the high wavelength tail of the principle LAB solvent emission (Figure~\ref{fig:LAB_Spectrum}), which can escape absorption but is depleted by the non-radiative coupling of LAB to the primary fluor (hence linking the fall time of this component with the rise time associated with the non-radiative transfer). The exception of pyrene might be explained by its greater absorption in this region. Nevertheless, we treat $\tau'$ here as a free phenomenological parameter in the fit. Finally, a delta function was used to represent the Cherenkov signal. The full optical response model is given in Equation~\ref{eq:optics}. The optical model was convolved with the system's IRF by asserting a [-5, 5] ns time cut about the IRF's peak and applying a discrete linear convolution to the two binned distributions. The result was then scaled to match the number of events in a relevant run and free model parameters were determined by minimizing the negative log likelihood.  The resulting fit model is given in Equation~\ref{eq:response} and the associated parameters are described in Table~\ref{tab:fit_parameters}. For each fluor, fits were performed simultaneously over both the \textit{towards} and \textit{away} spectra. The parameters $t_0$, $N_{events}$, $F_{Cheren}$ were allowed to float independently for each of the two spectra. All other parameters were assumed to be common to both. Quoted uncertainties were based on both values returned by the minimiser and, where possible, directly observed variations between multiple measurements.

\begin{equation}
    \centering
    f_{optics}(t) = (1 - F_{Cheren}) \left( \sum_i^n  \left( A_i\frac{e^{-\frac{t}{\tau_i}} - e^{-\frac{t}{\tau_{rise}}}} {\tau_i - \tau_{rise}}\right) + A'  \cdot \frac{e^{-\frac{t}{\tau'}}}{\tau'} \right) + F_{Cheren} \cdot \delta(t)
    \label{eq:optics}
\end{equation}

\begin{equation}
    \centering
    f_{response}(t) = N_{events} \cdot \left( f_{optics}(t - t_0) \circledast IRF(t)  \right)
    \label{eq:response}
\end{equation}

\vskip 0.3in

\begin{table}[H]
\begin{center}
\begin{tabular}{ c | c }
 Name & Description  \\ 
 \hline
$IRF$ & the system's impulse response function\\
$t_0$ & time offset (cable delays etc) \\
$\delta(t)$ & Dirac delta function \\
N$_{events}$ & number of events in this measurement \\
F$_{Cheren}$ & fractional contribution of Cherenkov light \\  
A$_i$ & fraction of scintillation light emitted in the ith component \\
$\tau_i$ & decay constant for the ith component \\
$\tau_{rise}$ & rise time of scintillator \\
A$'$ & fraction of sctintillation light emitted via residual LAB emission \\
$\tau'$ & fall time of residual LAB emission \\ 
\end{tabular}
\end{center}
\caption{Parameters in the scintillator time distribution fit}
\label{tab:fit_parameters}
\end{table}


\section{Acenaphthene}

\begin{figure}[H]
\centering
  \begin{subfigure}{0.8\textwidth}
    \centering
    \includegraphics[width=0.9\linewidth]{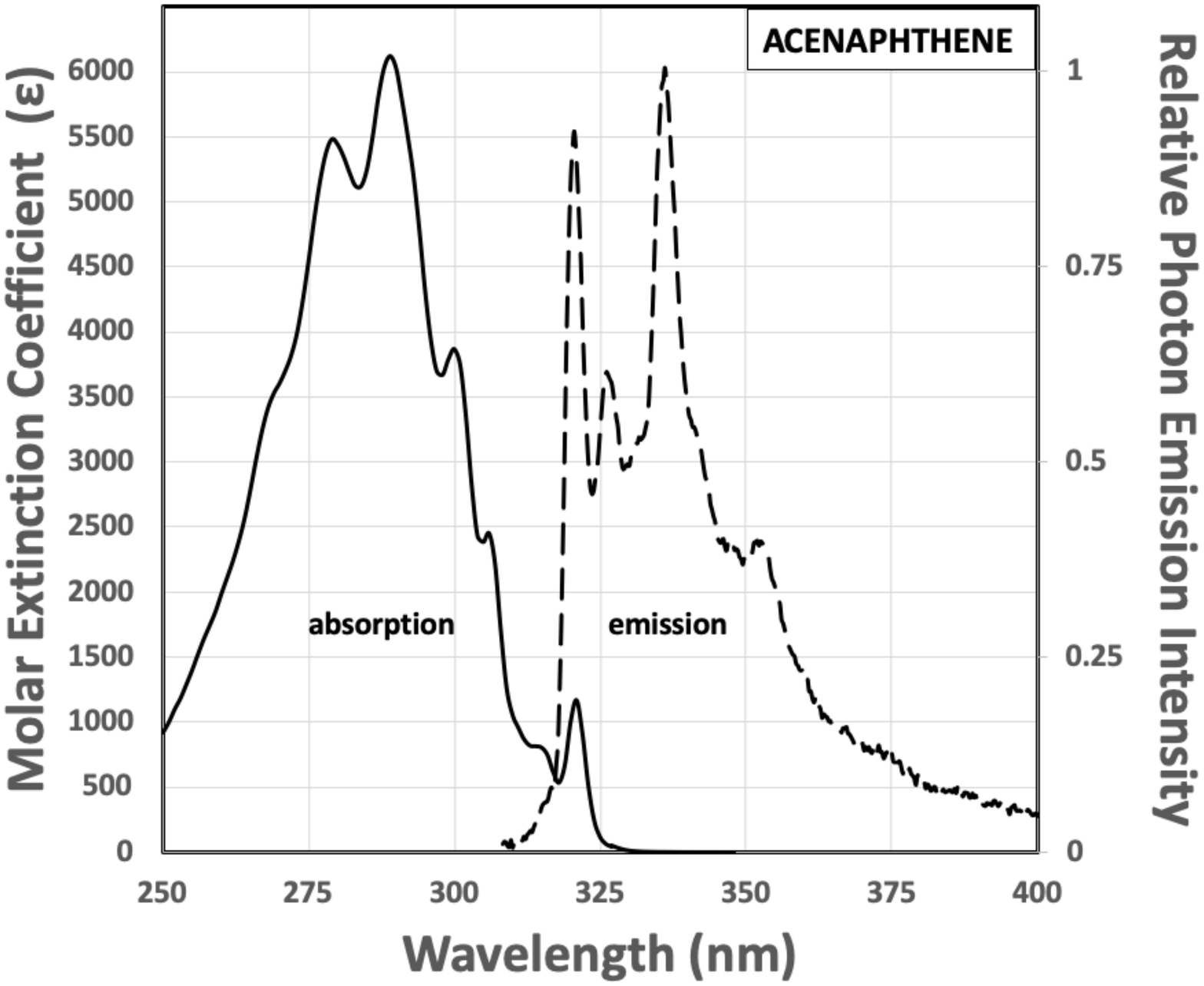}
    \caption{Absorption (in cyclohexane, solid line) and relative emission spectra (4g/L in LAB, dashed line)}
    \label{fig:Acenaphthene1}
  \end{subfigure} \\
  \vskip 0.3in
  \begin{subfigure}{0.8\textwidth}
    \centering
    \includegraphics[width=0.9\linewidth]{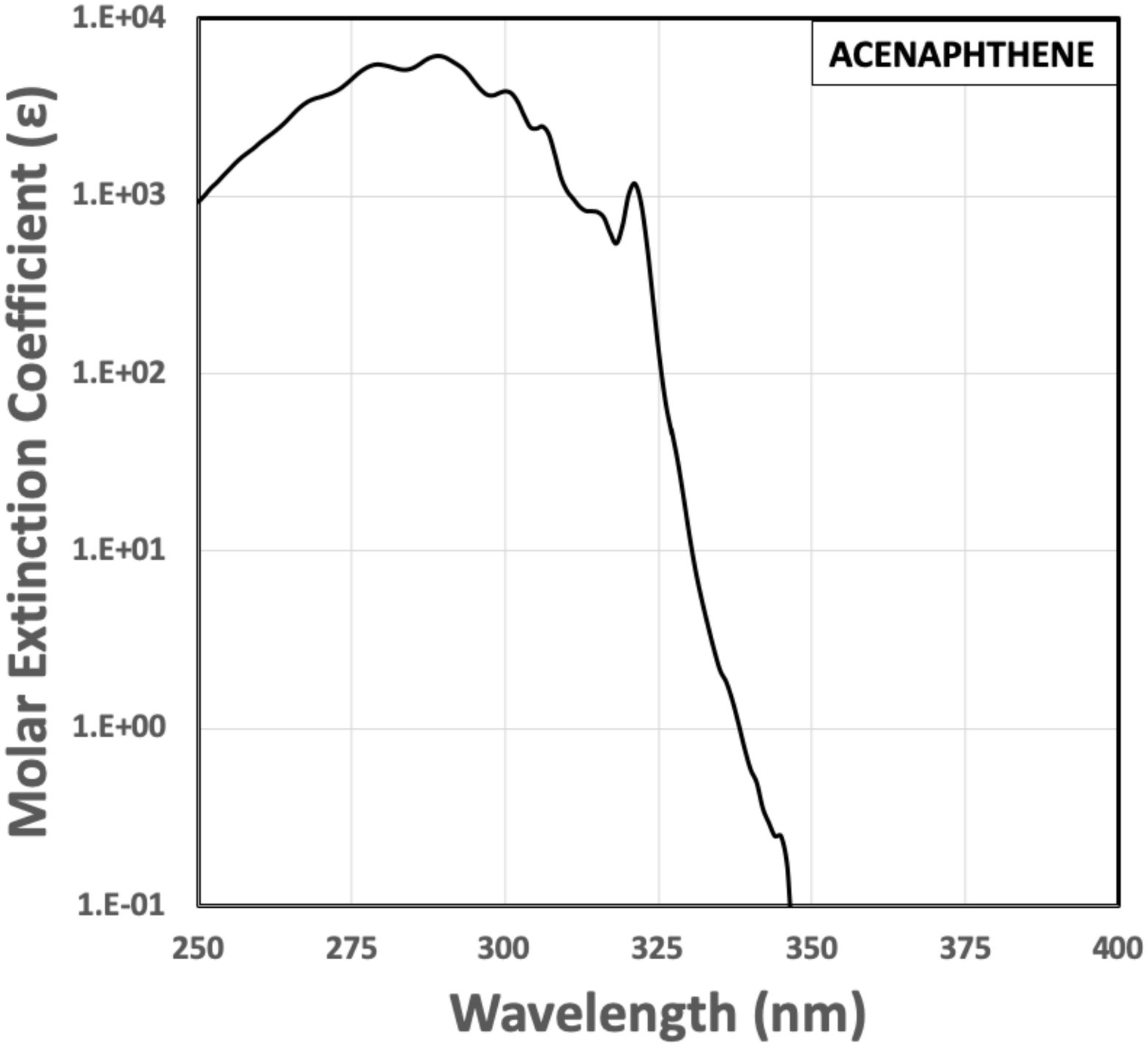}
    \caption{Logarithmic absorption spectrum}
    \label{fig:Acenaphthene2}
  \end{subfigure}
  \caption{Acenaphthene absorption spectrum in cyclohexane and relative emission spectrum in LAB.}
\end{figure}

Acenaphthene (CAS 83-32-9) is a colourless, needle-like crystalline solid with a melting point of 93$^o$C and a chemical formula of C$_{12}$H$_{10}$ (MW 154.212 g/mol) that comprises two fused benzene rings with an additional ethylene bridge. The acenaphthene sample used here was obtained from Tokyo Chemical Company (TCI) with $>$99\% purity. Figure \ref{fig:Acenaphthene1} shows the absorption and relative emission spectra in LAB, with figure \ref{fig:Acenaphthene2} showing more details of the absorption on a logarithmic scale.

When used as a primary fluor in LAB, the light yield was found to reach a maximum for a concentration near 4 g/l. This gave a light yield of $67\pm2$\% that of the PPO reference, which is just under what would be expected from the respective quantum yields typically quoted (0.6 for acenaphthene \cite{Berlman} and $\sim$0.8-0.84 for PPO \cite{PPO}, \cite{PPO2}) assuming similar coupling efficiencies. Owing to the light emission peaking near $\sim$335nm, in some applications it may be prudent to use acenaphthene in conjunction with a secondary fluor, such as bis-MSB, to shift the emission wavelength beyond the LAB absorption region for large detectors.

\begin{figure}[H]
  \begin{subfigure}{0.5\textwidth}
    \centering
    \includegraphics[width=0.9\linewidth]{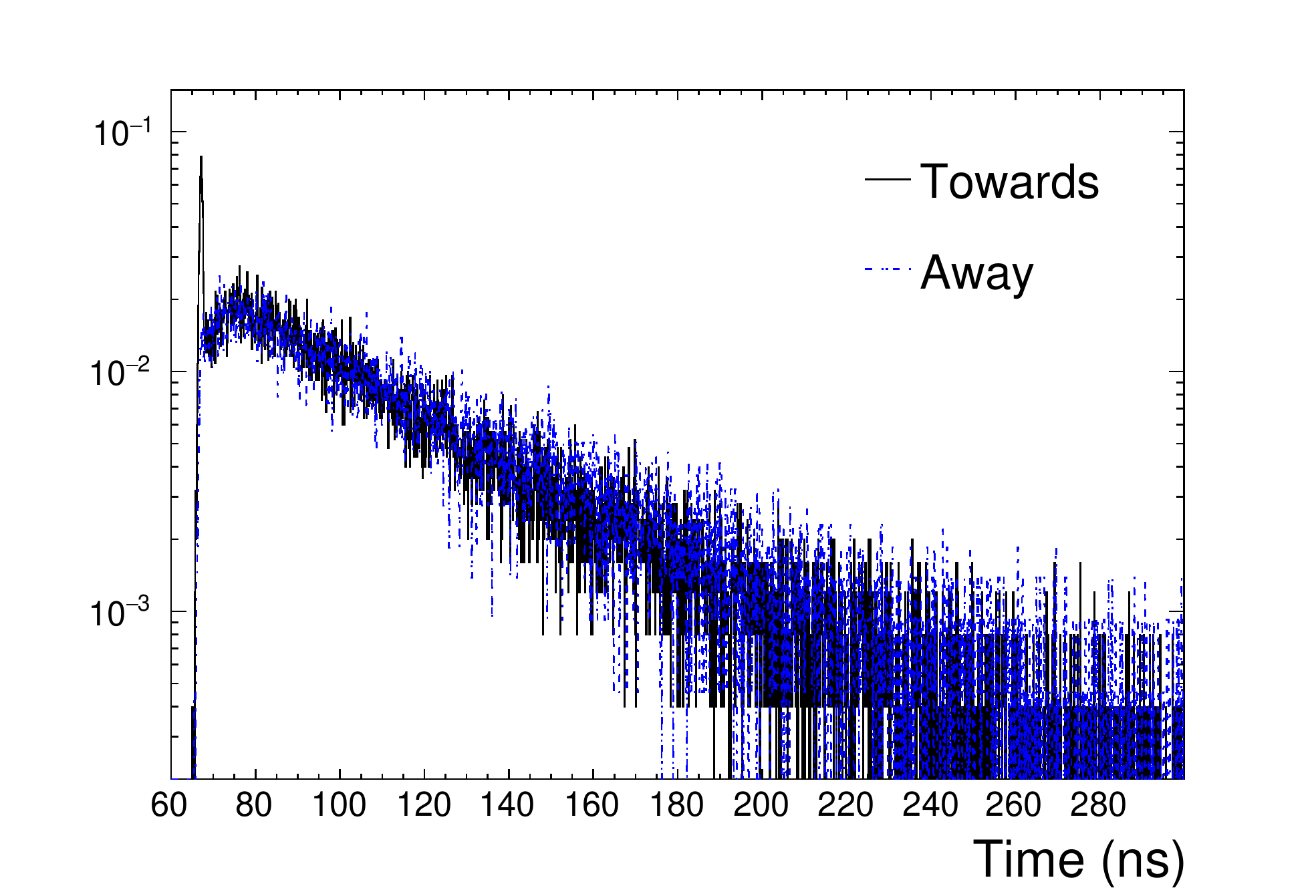}
    \caption{timing spectra (\textit{time measurement PMT})}
    \label{fig:compare_time_acenapthene}
  \end{subfigure}%
  \begin{subfigure}{0.5\textwidth}
    \centering
    \includegraphics[width=0.9\linewidth]{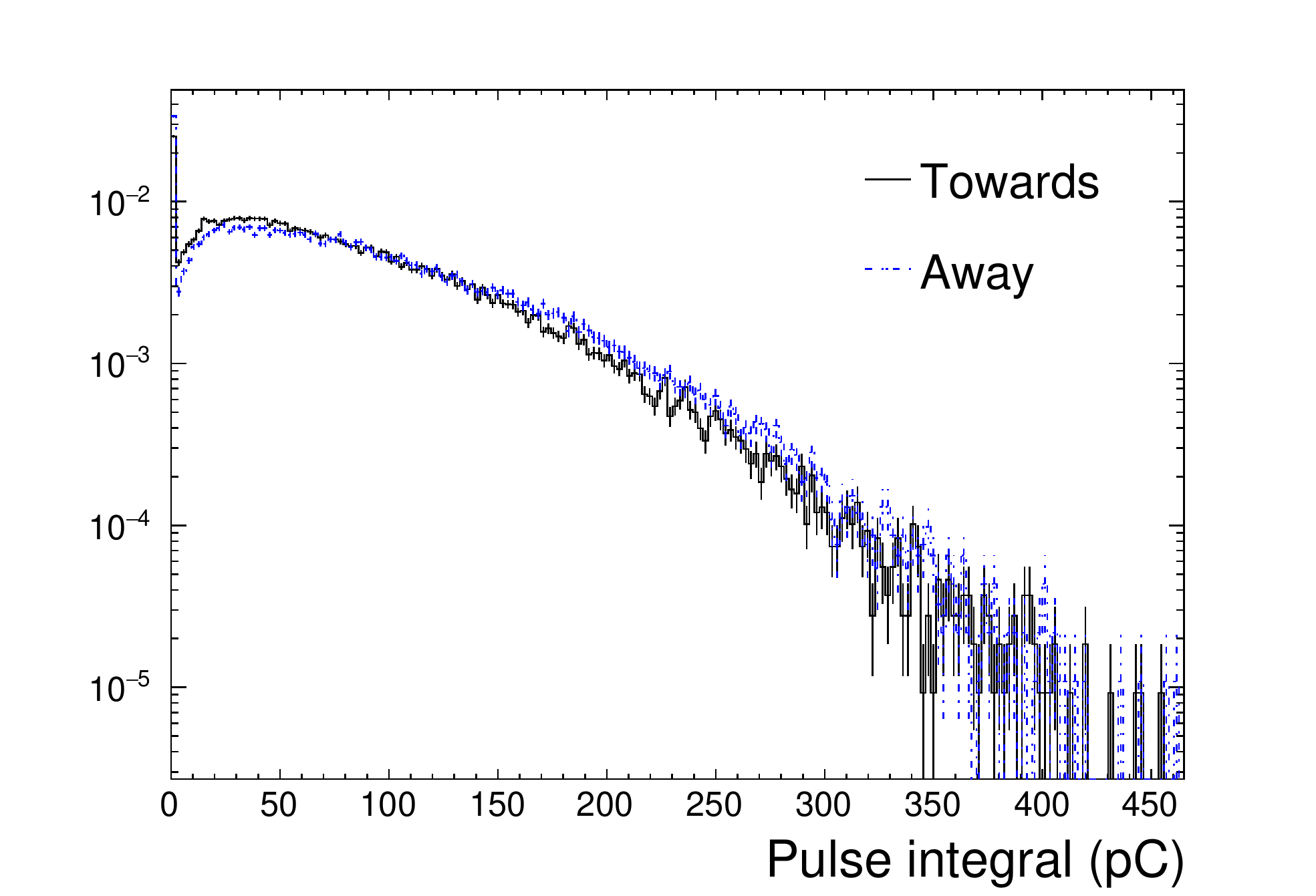}
    \caption{charge spectra (\textit{charge collection PMT})} 
    \label{fig:compare_charge_acenapthene}
  \end{subfigure}
  \vskip\baselineskip
  \begin{subfigure}{0.5\textwidth}
    \centering
    \includegraphics[width=0.9\linewidth]{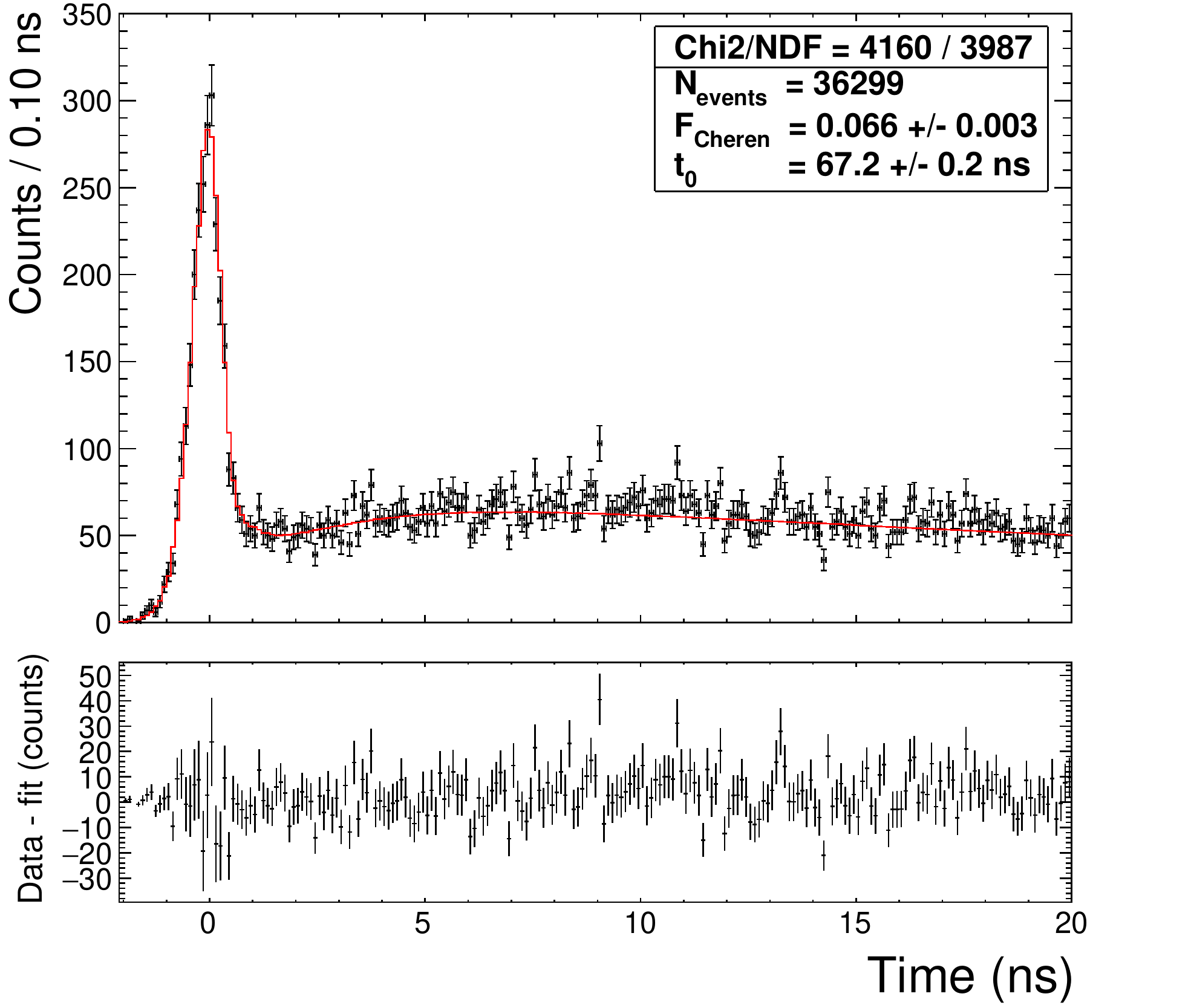}
    \caption{fit to \textit{towards} configuration}
    \label{fig:fit_acenapthene_towards}
  \end{subfigure}%
  \begin{subfigure}{0.5\textwidth}
    \centering
    \includegraphics[width=0.9\linewidth]{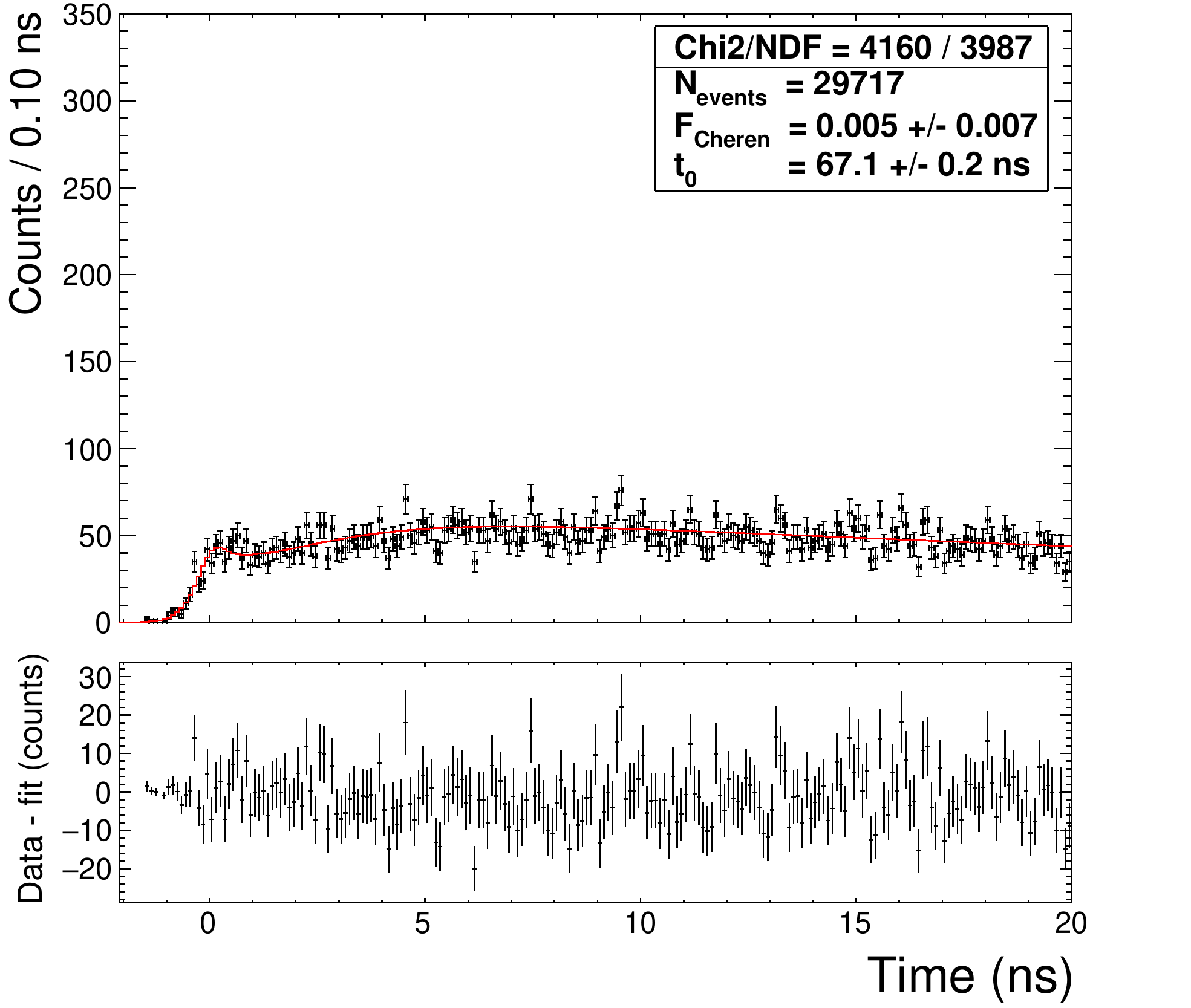}
    \caption{fit to \textit{away} configuration}
    \label{fig:fit_acenapthene_away}
  \end{subfigure}
  \caption{Time profile results for acenapthene. The fit parameters associated with the scintillation light are given in Tables~\ref{tab:time_constants} and \ref{tab:scale_constants}.}
\end{figure}

The timing spectra for the forward and backward experimental configurations are given in Figure~\ref{fig:compare_time_acenapthene}. The results of fits to the measured timing spectra are given in Tables~\ref{tab:time_constants} and \ref{tab:scale_constants}, showing a rise time of $2.1\pm0.2$ ns and a decay time of $45.4\pm0.3$ ns. Measurements of the primary decay time component were found to be comparable with measurements of acenepthene in cyclohexane by \cite{Berlman}. This can be compared, for example, to the approach of \cite{Guo} using dilute PPO concentrations, which achieved a scintillator formulation with a rise time of 1.16~ns, a decay time of 26.7~ns and a relative light yield corresponding to $\sim$35\% that of more standard PPO formulations ($\sim$2-3 g/L). By contrast, the acenaphthene scintillator has nearly double the light yield with much better time separation. Note that the Cherenkov separation shown in Figures~\ref{fig:fit_acenapthene_towards} and \ref{fig:fit_acenapthene_away} is substantial, with a very clear directional peak with little contamination by scintillation light at early times. 

\section{Pyrene}

\begin{figure}[H]
\centering
  \begin{subfigure}{0.7\textwidth}
    \centering
    \includegraphics[width=0.9\linewidth]{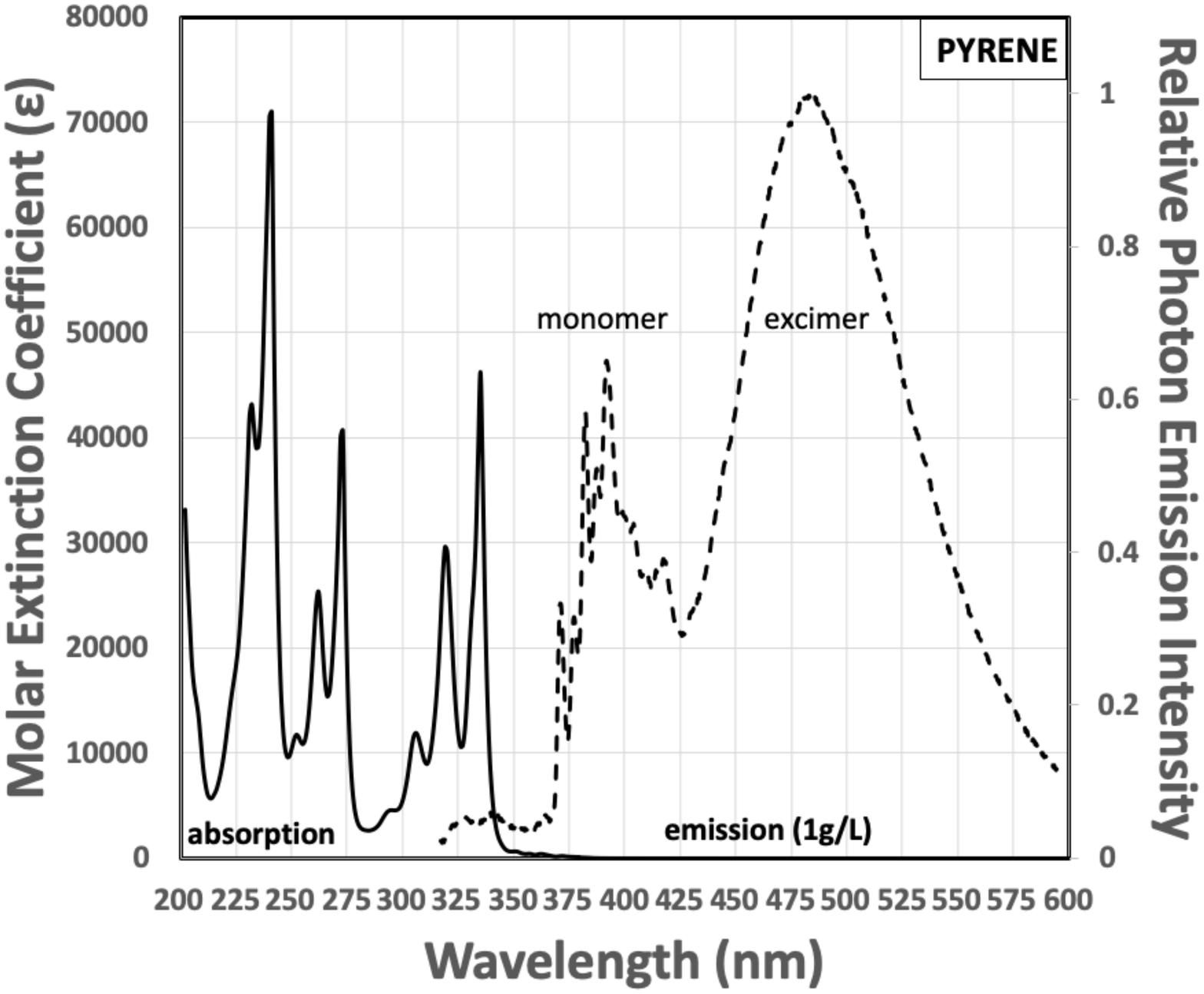}
    \caption{Absorption (in cyclohexane, solid line) and relative emission spectra (1g/L in LAB, dashed line)}
    \label{fig:Pyrene1}
  \end{subfigure} \\
  \vskip 0.3in
  \begin{subfigure}{0.7\textwidth}
    \centering
    \includegraphics[width=0.9\linewidth]{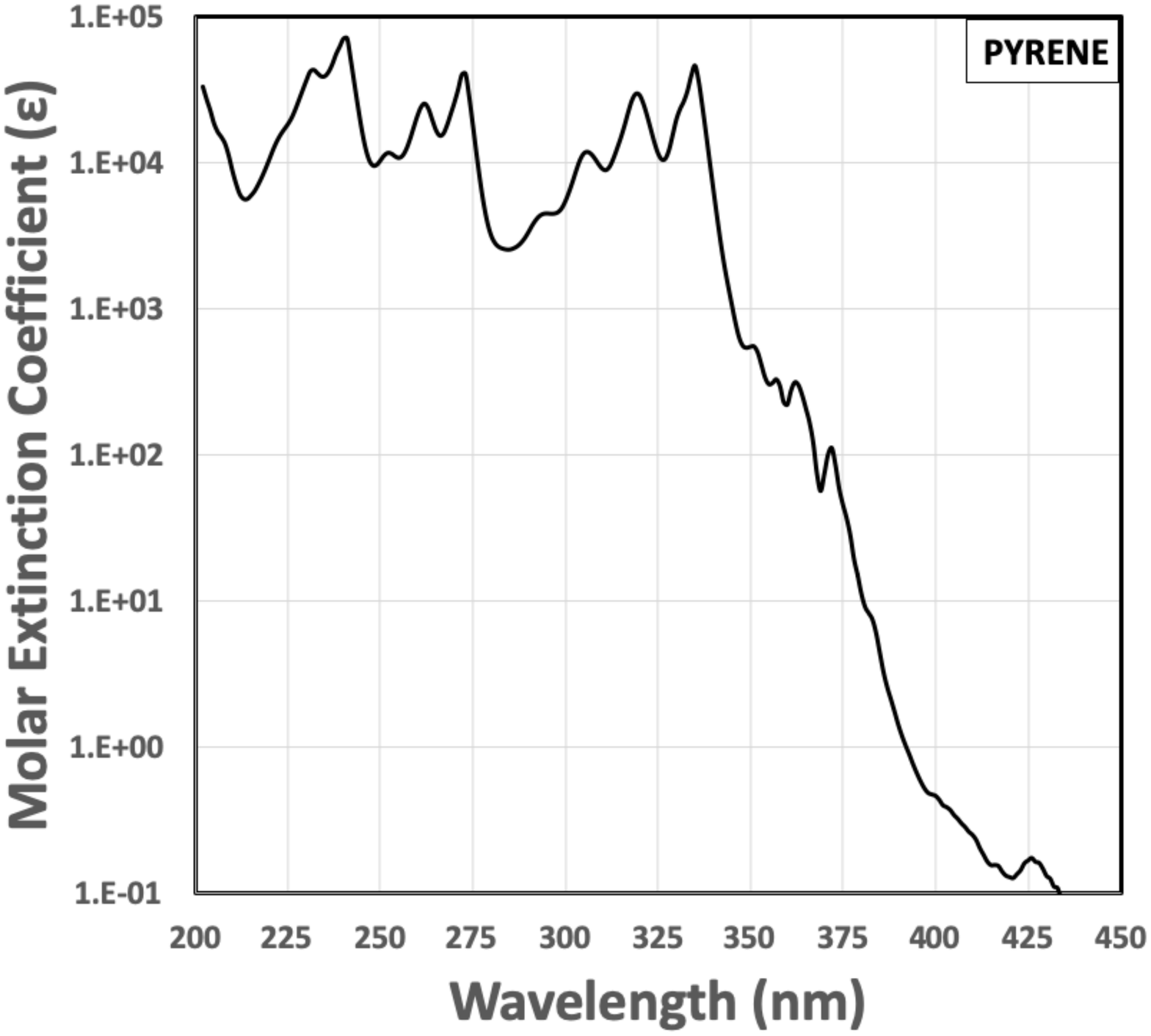}
    \caption{Logarithmic absorption spectrum}
    \label{fig:Pyrene2}
  \end{subfigure}
  \caption{Pyrene absorption spectrum in cyclohexane and relative emission spectrum in LAB.}
\end{figure}

Pyrene (CAS 129-00-0) is a pale yellow, crystalline solid with a melting point of 150$^o$C and a chemical formula of C$_{16}$H$_{10}$ (MW 202.25~g/mol) that comprises 4 fused benzene rings. The pyrene sample used here was obtained from Sigma Aldrich (Merck) with $>$99~\% purity. Figure \ref{fig:Pyrene1} shows the absorption and relative emission spectra in LAB, with figure \ref{fig:Pyrene2} showing more details of the absorption on a logarithmic scale.

Pyrene exhibits higher wavelength excimer emission, peaking around 480 nm, which becomes more prominent at higher concentrations and is sensitive to the solvent used. The emission shape shown in figure \ref{fig:Pyrene1} is for a concentration of 1~g/l in PPO. The leftmost monomer peak, centered around $\sim$390~nm, becomes negligible for concentrations of several g/l or more. 

When used as a primary fluor in LAB, the light yield was found to reach a maximum for concentrations beyond 2~g/l, maintaining an approximately constant level for concentrations up to at least 10~g/l. This is a consequence of the high-yield excimer emission at higher wavelengths that largely avoids self-absorption. As an aside, we note that the quantum yield for pyrene of 0.32 quoted by Berlman \cite{Berlman} appears to be incorrect and more recent measurements (for example \cite{Katoh}) appear to confirm the earlier measurements by Medinger {\em et al.} \cite{pyrene} of closer to 0.65.  The light yield at a concentration of 1~g/l corresponds to $76\pm2$~\% that of the PPO reference. This is consistent with a pyrene quantum yield of $\sim$0.65 compared to $\sim$0.8 for PPO, particularly considering that the coupling efficiency might be slightly lower compared with the higher concentration PPO reference. For pyrene concentrations in excess of several g/l, this light yield rises to $99\pm6$~\%, which would be consistent with a higher coupling efficiency. It should be noted that, as a consequence of the excimer emission occurring at higher wavelengths that are away from the peak of bialkalai photocathode efficiencies, the observed light levels tend to be $\sim$30\% lower for typical large-format PMTs. On the other hand, absorption in LAB is reduced at these wavelengths, which may compensate for light levels to some extent in large scale detectors. 

The timing spectra for the forward and backward experimental configurations of 
the monomer state, selected with a 400~nm short pass optical filter, at a concentration of 1~g/l are shown in Figure~8. 
Spectra for the excimer state, selected with a 450~nm long pass optical filter, at concentrations of 1~g/l and 8~g/l are shown in Figures~9 
and 10, 
respectively. The results of fits to the measured timing spectra are given in Tables~\ref{tab:time_constants} and \ref{tab:scale_constants}, showing rise times ranging from $\sim$4.5-60 ns and decay times ranging from $\sim$50-100 ns, depending on concentratons used. The Cherenkov separation is even more distinct than for acenaphthene, with a very clear directional peak with little contamination by scintillation light at early times. We note that there is a more prominent small bump of Cherenkov light in the backwards direction compared with acenaphthene. We believe this is due to a lower energy threshold for this configuration as a result of more scintillation light (note that the fraction of Cherenkov light is smaller). As lower energy electrons are more easily deflected by multiple scattering, the angular distribution of Cherenkov light will therefore be broadened (though we have not tried to model this quantitatively).

For the concentrations used here, the A$'$ component fit to a very small values, often consistent with zero. We believe this is consistent with the higher molar extinction at longer wavelength for pyrene compared with primary fluors such as acenaphthene or PPO, which then absorbs more of the higher wavelength residual LAB emission. Similarly, DPA and DPH are considered here as secondary fluors at much lower concentrations and, hence, also with much less absorption in the higher wavelength region than pyrene. Measurements of the primary fall time components of both the monomer and excimer states, shown in Figure~\ref{fig:pyrene_comparison_summary}, were found to be comparable with measurements of pyrene in cyclohexane as measured by \cite{Birks}, though there are clear differences in the measured rise times, which may be indicative of solvent effects.

\begin{figure}[H]
  \begin{subfigure}{0.5\textwidth}
    \centering
    \includegraphics[width=0.9\linewidth]{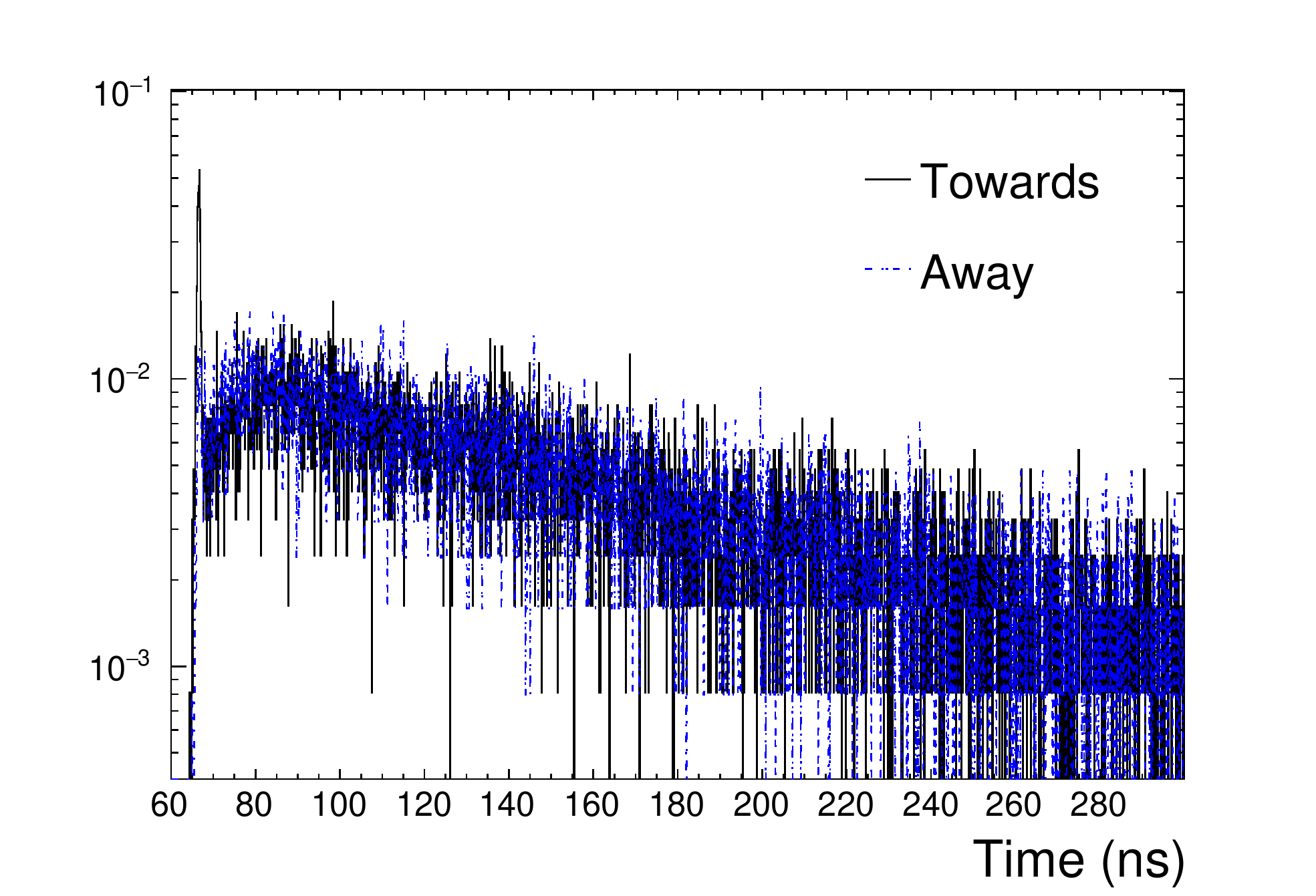}
    \caption{timing spectra (\textit{time measurement PMT})}
    \label{fig:compare_time_pyrene_monomer}
  \end{subfigure}%
  \begin{subfigure}{0.5\textwidth}
    \centering
    \includegraphics[width=0.9\linewidth]{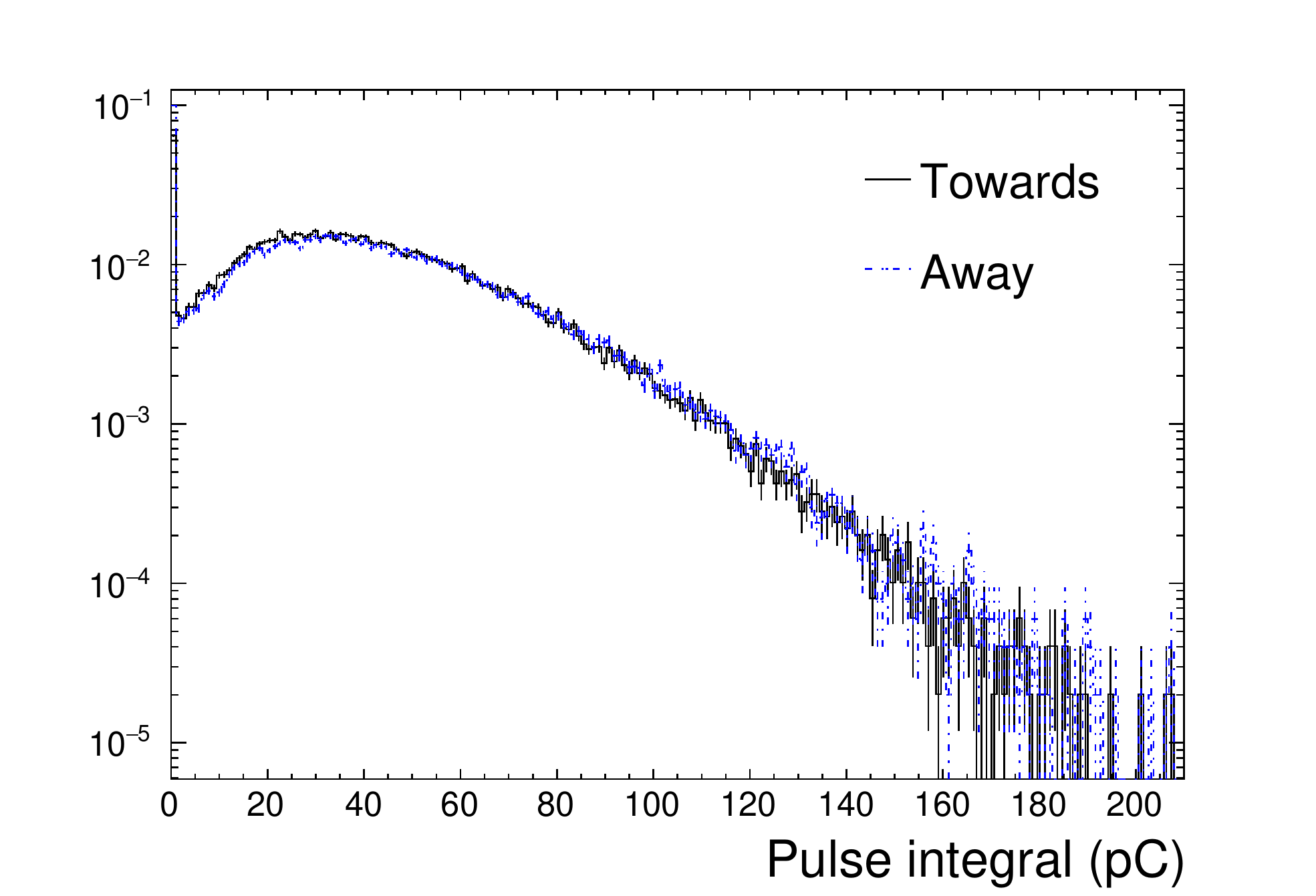}
    \caption{charge spectra (\textit{charge collection PMT})} 
    \label{fig:compare_charge_pyrene_monomer}
  \end{subfigure}
  \vskip\baselineskip
  \begin{subfigure}{0.5\textwidth}
    \centering
    \includegraphics[width=0.9\linewidth]{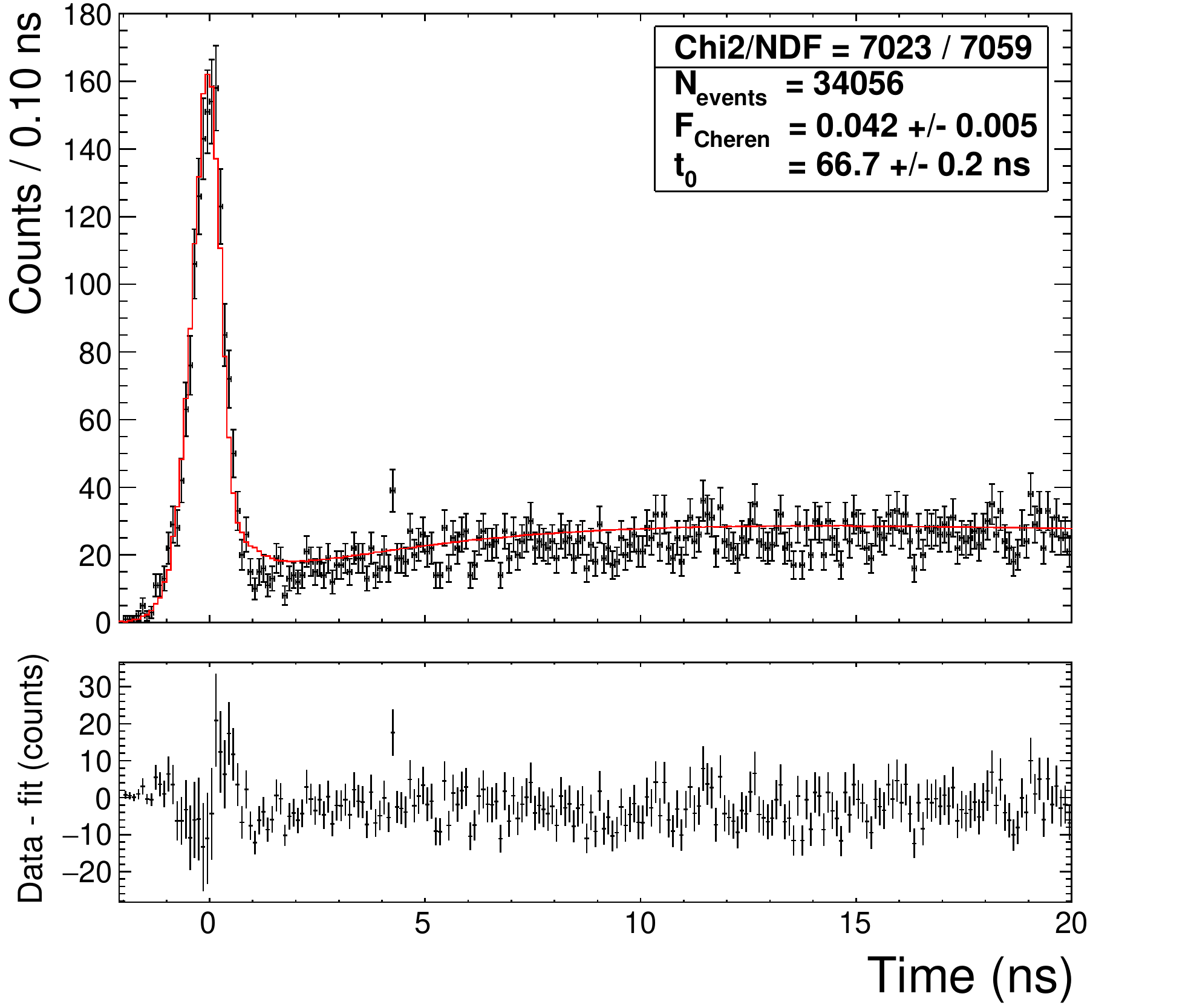}
    \caption{fit to \textit{towards} configuration}
    \label{fig:fit_pyrene_monomer_towards}
  \end{subfigure}%
  \begin{subfigure}{0.5\textwidth}
    \centering
    \includegraphics[width=0.9\linewidth]{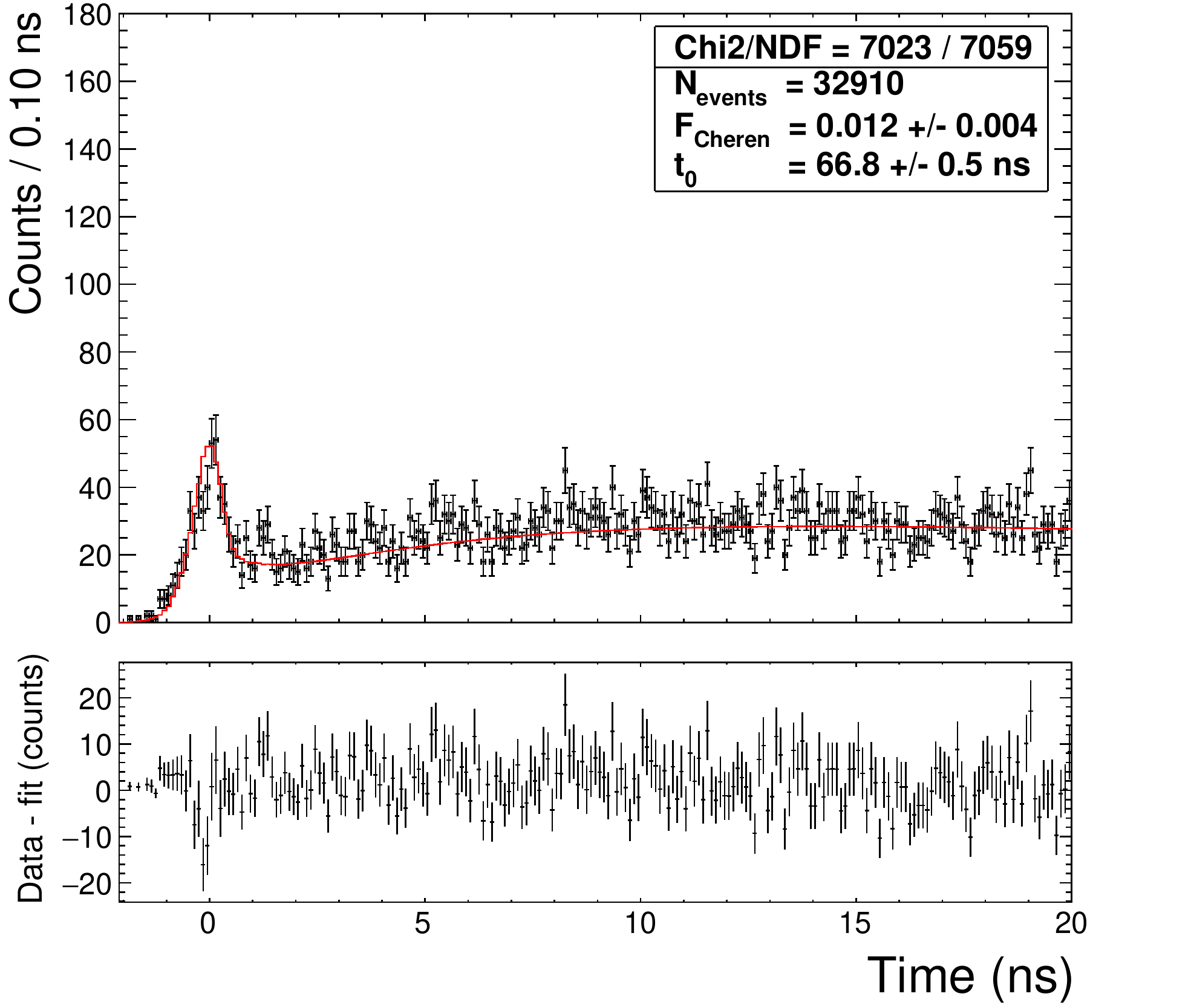}
    \caption{fit to \textit{away} configuration}
    \label{fig:fit_pyrene_monomer_away}
  \end{subfigure}
  \label{fig:pyrene_monomer}
  \caption{Time profile results for 1 g/l Pyrene with a 400 nm short pass filter, selecting the monomer state. The fit parameters associated with the scintillation light are given in Tables~\ref{tab:time_constants} and \ref{tab:scale_constants}.}
\end{figure}

\begin{figure}[H]
  \begin{subfigure}{0.5\textwidth}
    \centering
    \includegraphics[width=0.9\linewidth]{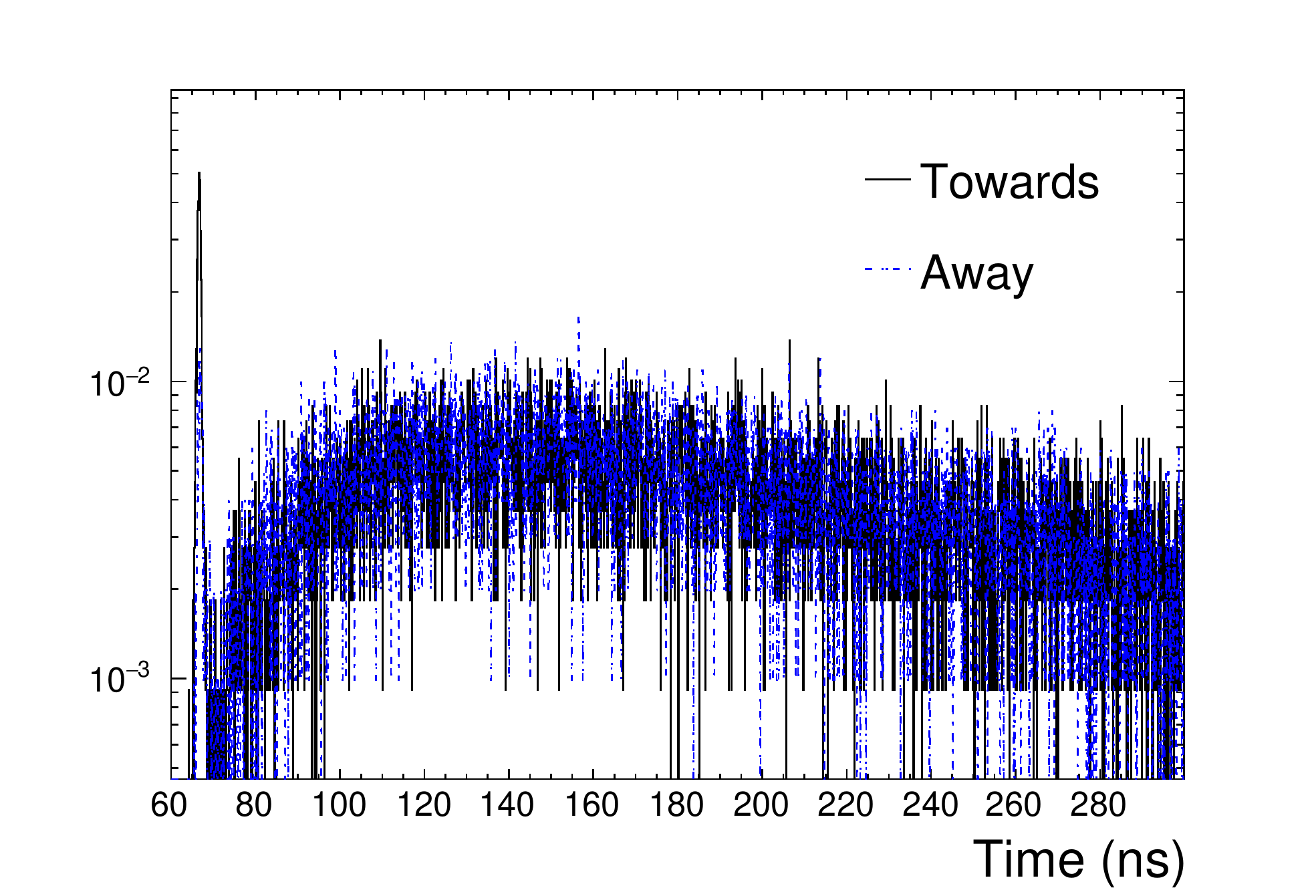}
    \caption{timing spectra (\textit{time measurement PMT})}
    \label{fig:compare_time_pyrene_1gl}
  \end{subfigure}%
  \begin{subfigure}{0.5\textwidth}
    \centering
    \includegraphics[width=0.9\linewidth]{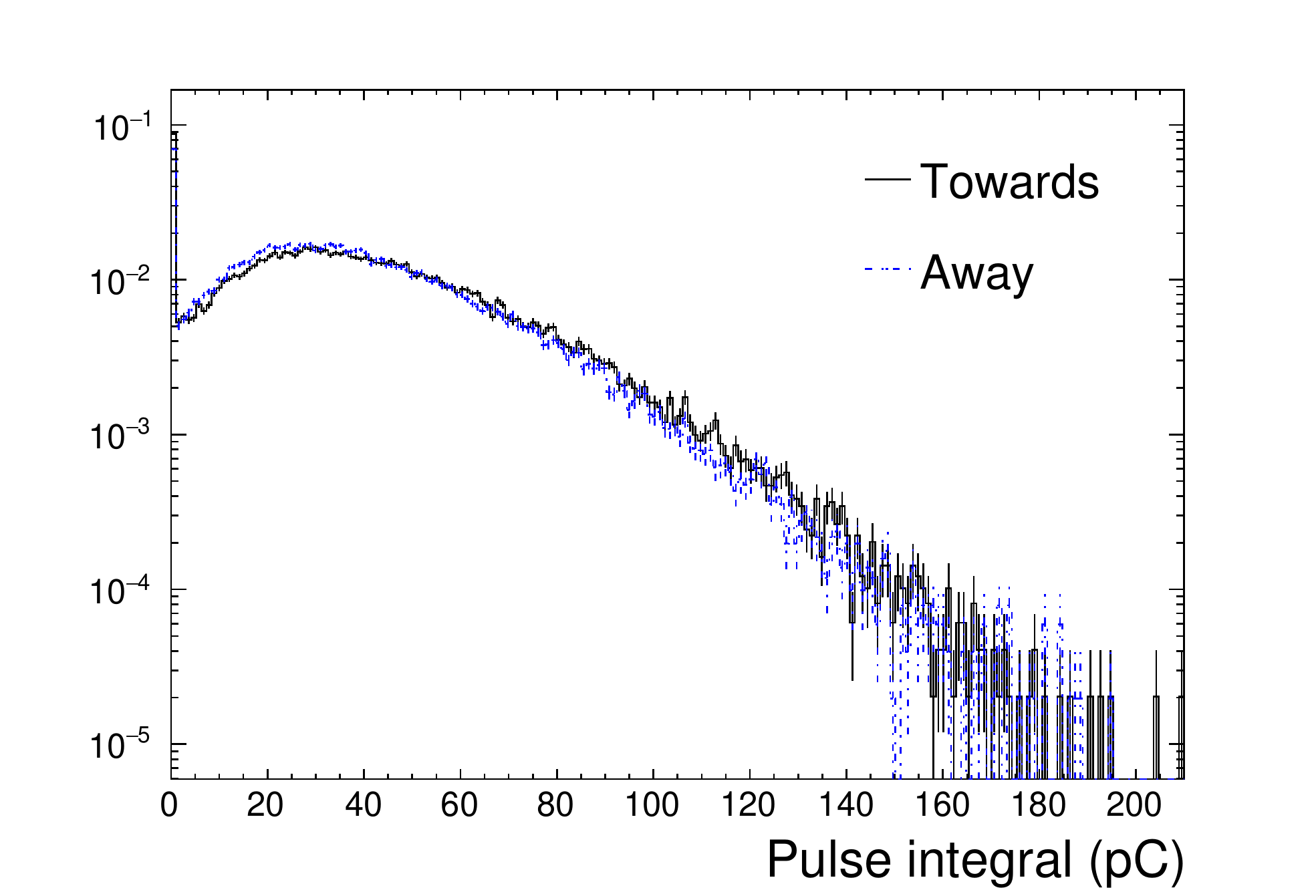}
    \caption{charge spectra (\textit{charge collection PMT})} 
    \label{fig:compare_charge_pyrene_eximer_1gl}
  \end{subfigure}
  \vskip\baselineskip
  \begin{subfigure}{0.5\textwidth}
    \centering
    \includegraphics[width=0.9\linewidth]{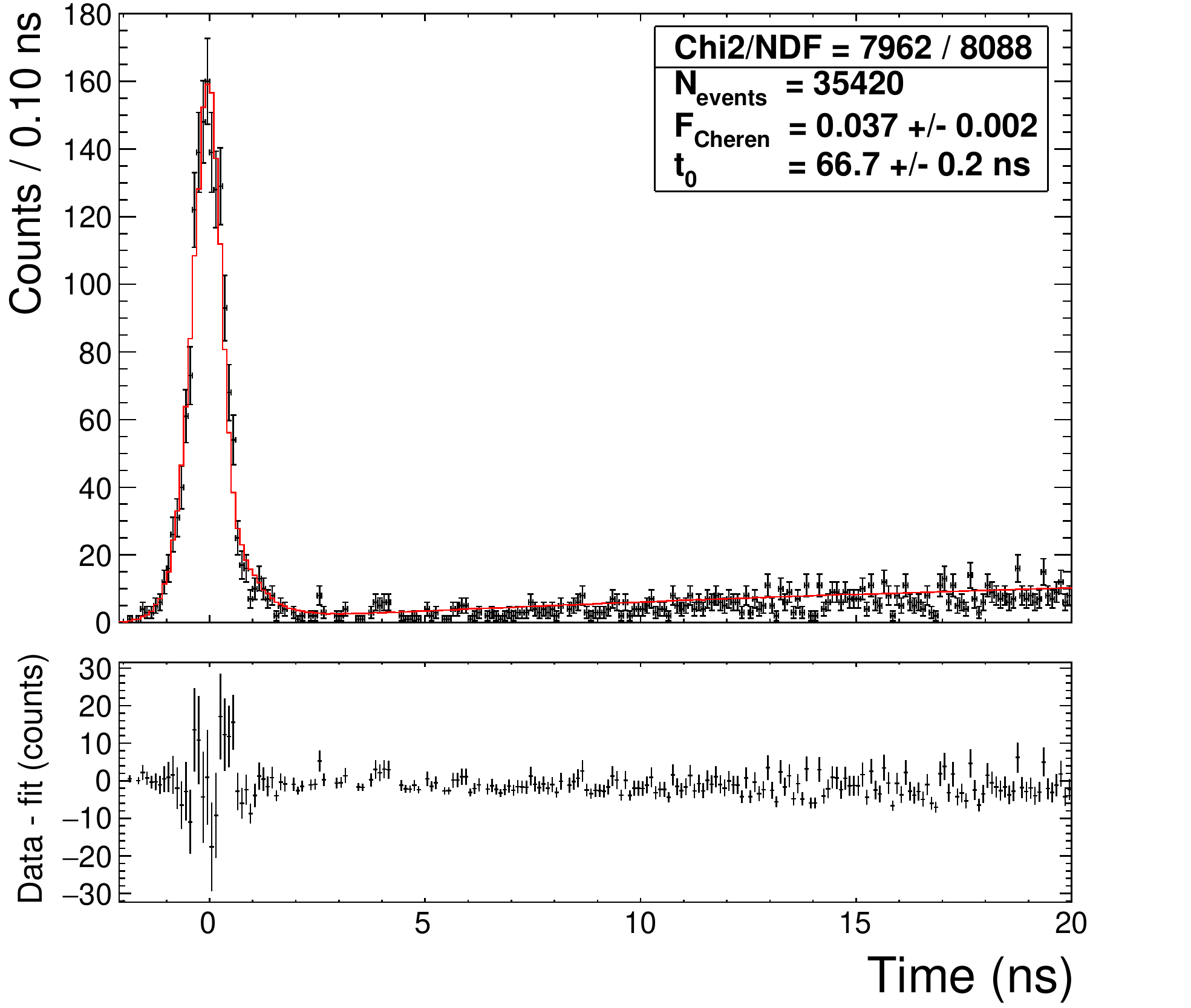}
    \caption{fit to \textit{towards} configuration}
    \label{fig:fit_pyrene_eximer_1gl_towards}
  \end{subfigure}%
  \begin{subfigure}{0.5\textwidth}
    \centering
    \includegraphics[width=0.9\linewidth]{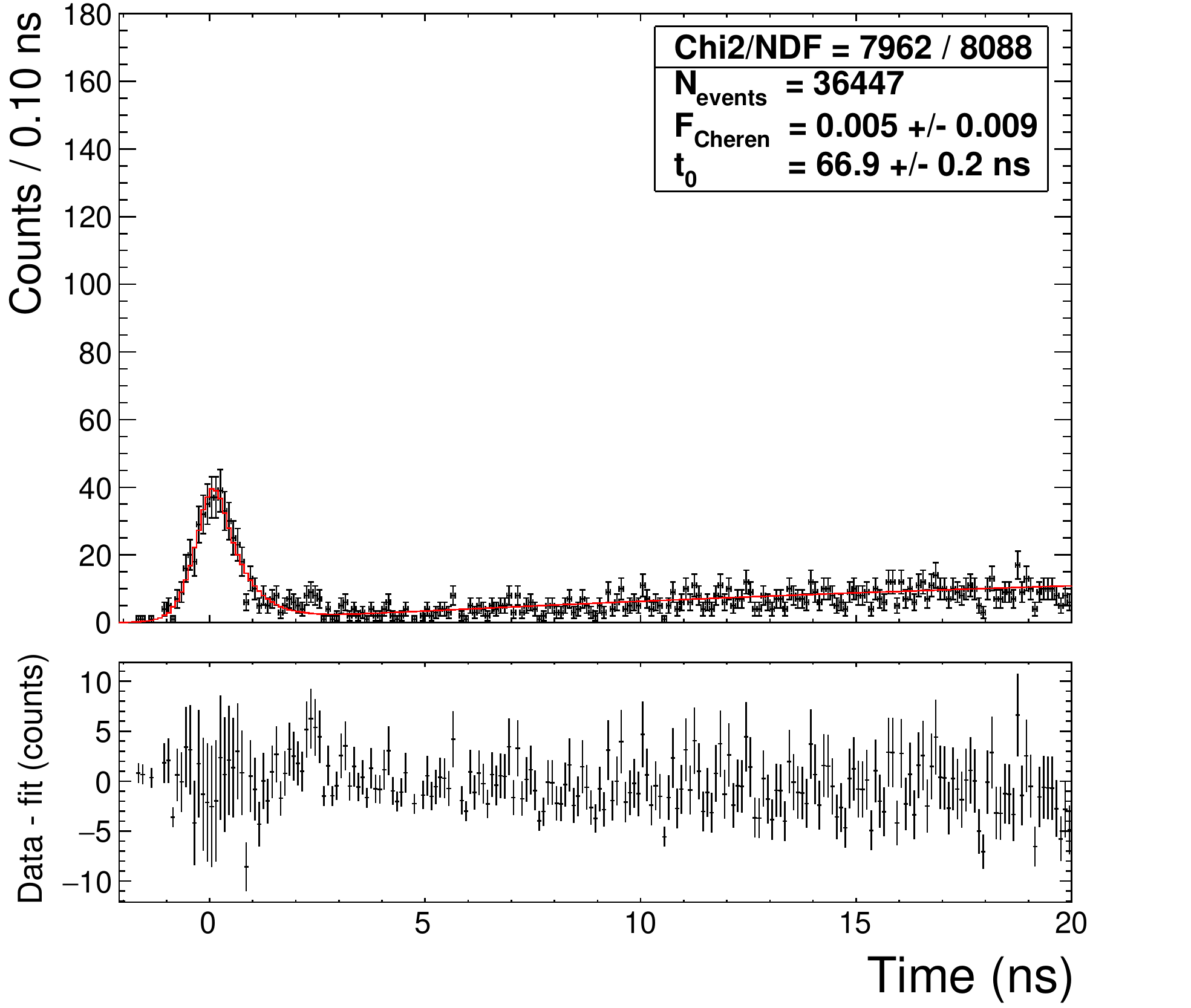}
    \caption{fit to \textit{away} configuration}
    \label{fig:fit_pyrene_eximer_1gl_away}
  \end{subfigure}
  \label{fig:pyrene_excimer_1gl}
  \caption{Time profile results for 1~g/l Pyrene with a 450~nm long pass filter, selecting the excimer state. The fit parameters associated with the scintillation light are given in Tables~\ref{tab:time_constants} and \ref{tab:scale_constants}.}
\end{figure}

\begin{figure}[H]
  \begin{subfigure}{0.5\textwidth}
    \centering
    \includegraphics[width=0.9\linewidth]{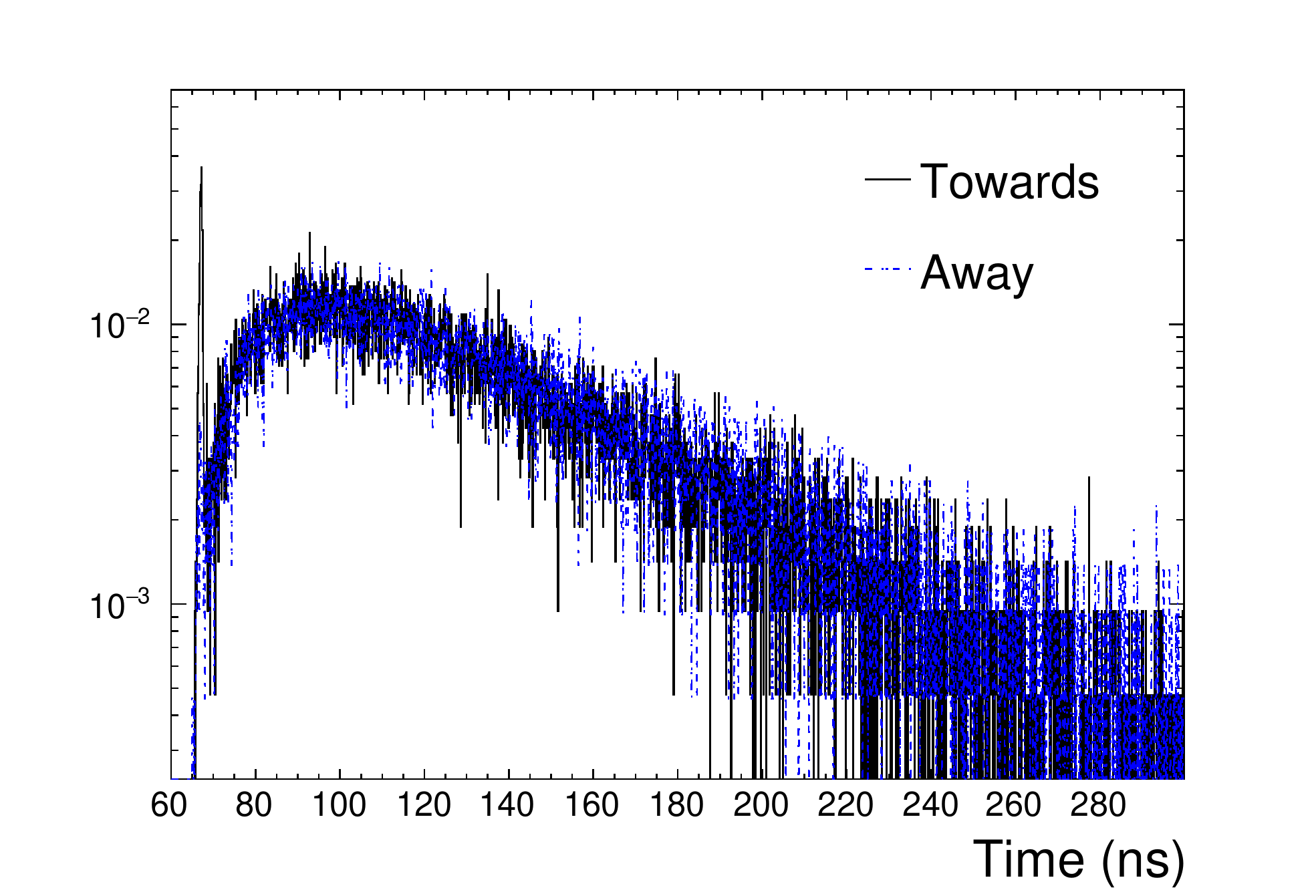}
    \caption{timing spectra (\textit{time measurement PMT})}
    \label{fig:compare_time_pyrene_eximer_10gl}
  \end{subfigure}%
  \begin{subfigure}{0.5\textwidth}
    \centering
    \includegraphics[width=0.9\linewidth]{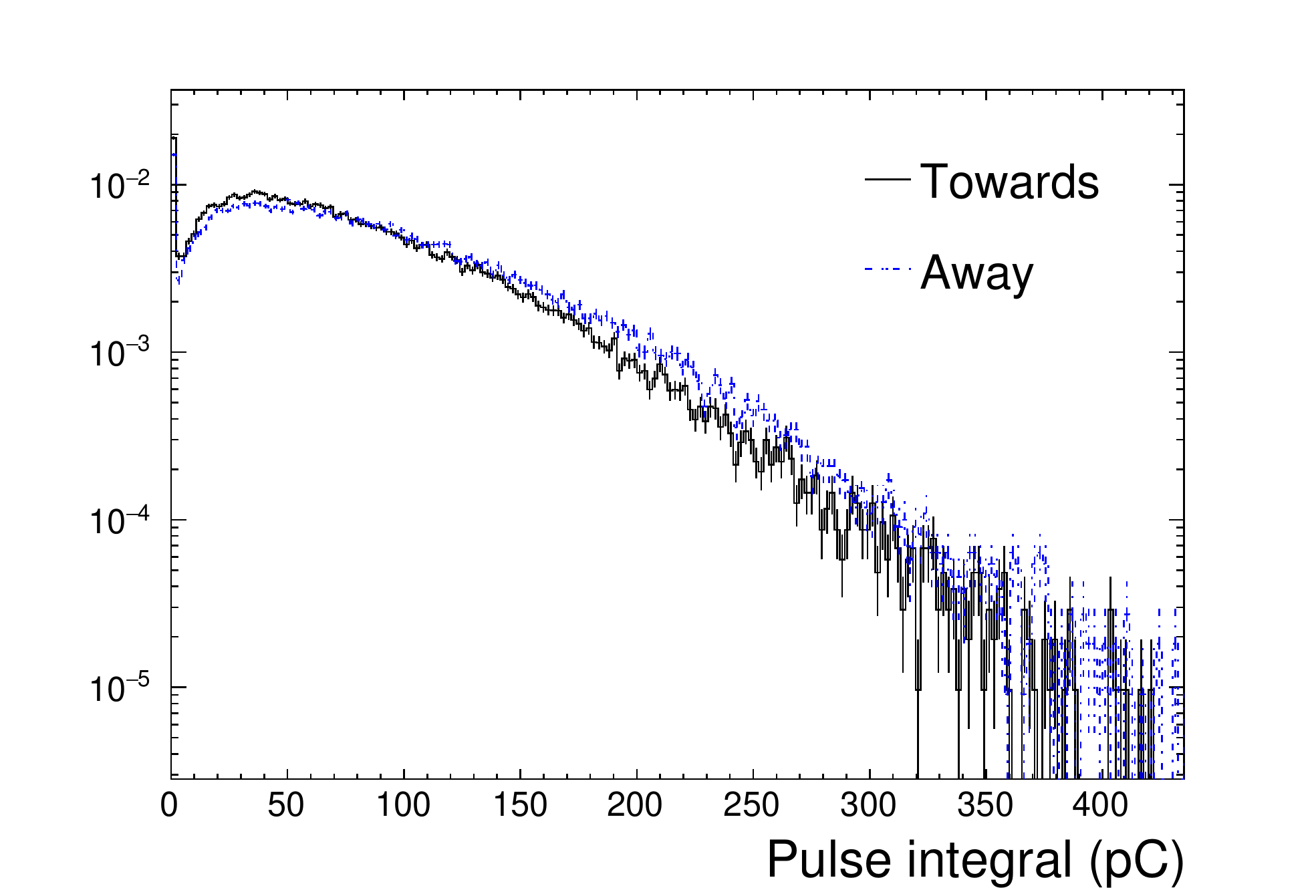}
    \caption{charge spectra (\textit{charge collection PMT})} 
    \label{fig:compare_charge_pyrene_eximer_10gl}
  \end{subfigure}
  \vskip\baselineskip
  \begin{subfigure}{0.5\textwidth}
    \centering
    \includegraphics[width=0.9\linewidth]{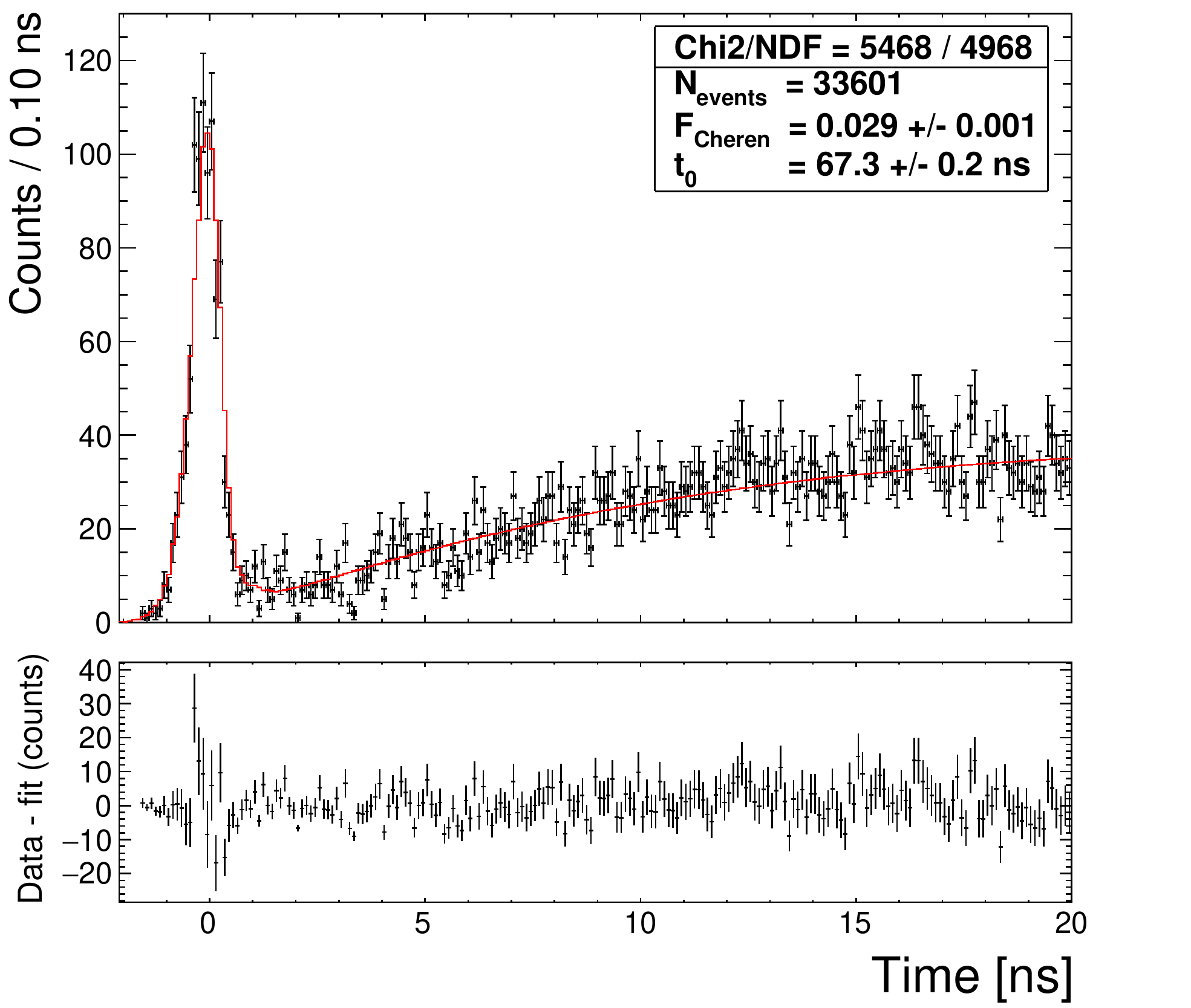}
    \caption{fit to \textit{towards} configuration}
    \label{fig:fit_pyrene_eximer_10gl_towards}
  \end{subfigure}%
  \begin{subfigure}{0.5\textwidth}
    \centering
    \includegraphics[width=0.9\linewidth]{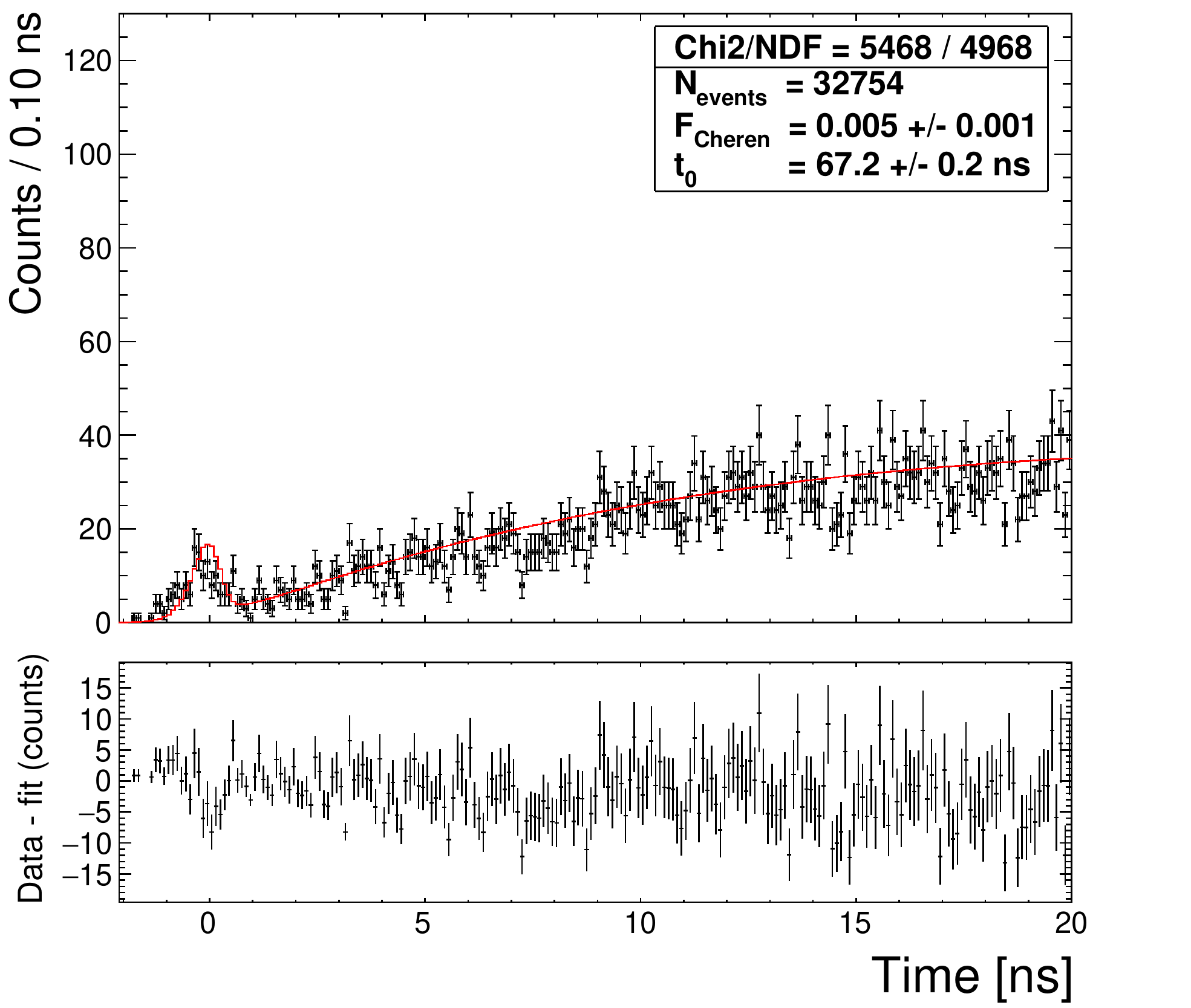}
    \caption{fit to \textit{away} configuration}
    \label{fig:fit_pyrene_eximer_10gl_away}
  \end{subfigure}
  \label{fig:pyrene_excimer_10gl}
  \caption{Time profile results for 8~g/l Pyrene with a 450~nm long pass filter, selecting the excimer state. The fit parameters associated with the scintillation light are given in Tables~\ref{tab:time_constants} and \ref{tab:scale_constants}.}
\end{figure}

\begin{figure}[H]
\centering
  \begin{subfigure}{0.9\textwidth}
    \centering
    \includegraphics[width=0.9\linewidth]{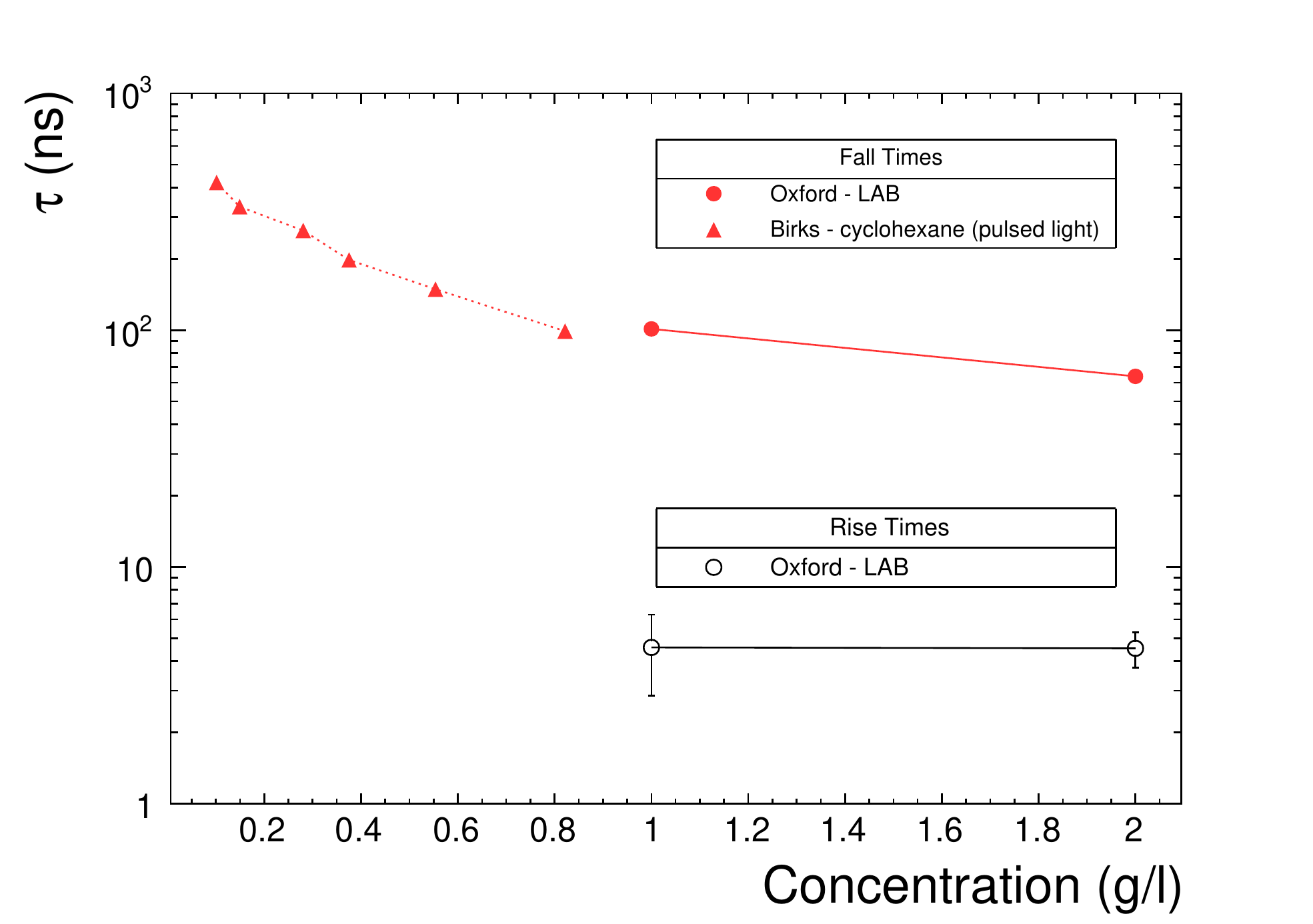}
    \caption{Monomer state}
    \label{fig:pyrene_monomer_summary}
  \end{subfigure}%
  \newline
  \begin{subfigure}{0.9\textwidth}
    \centering
    \includegraphics[width=0.9\linewidth]{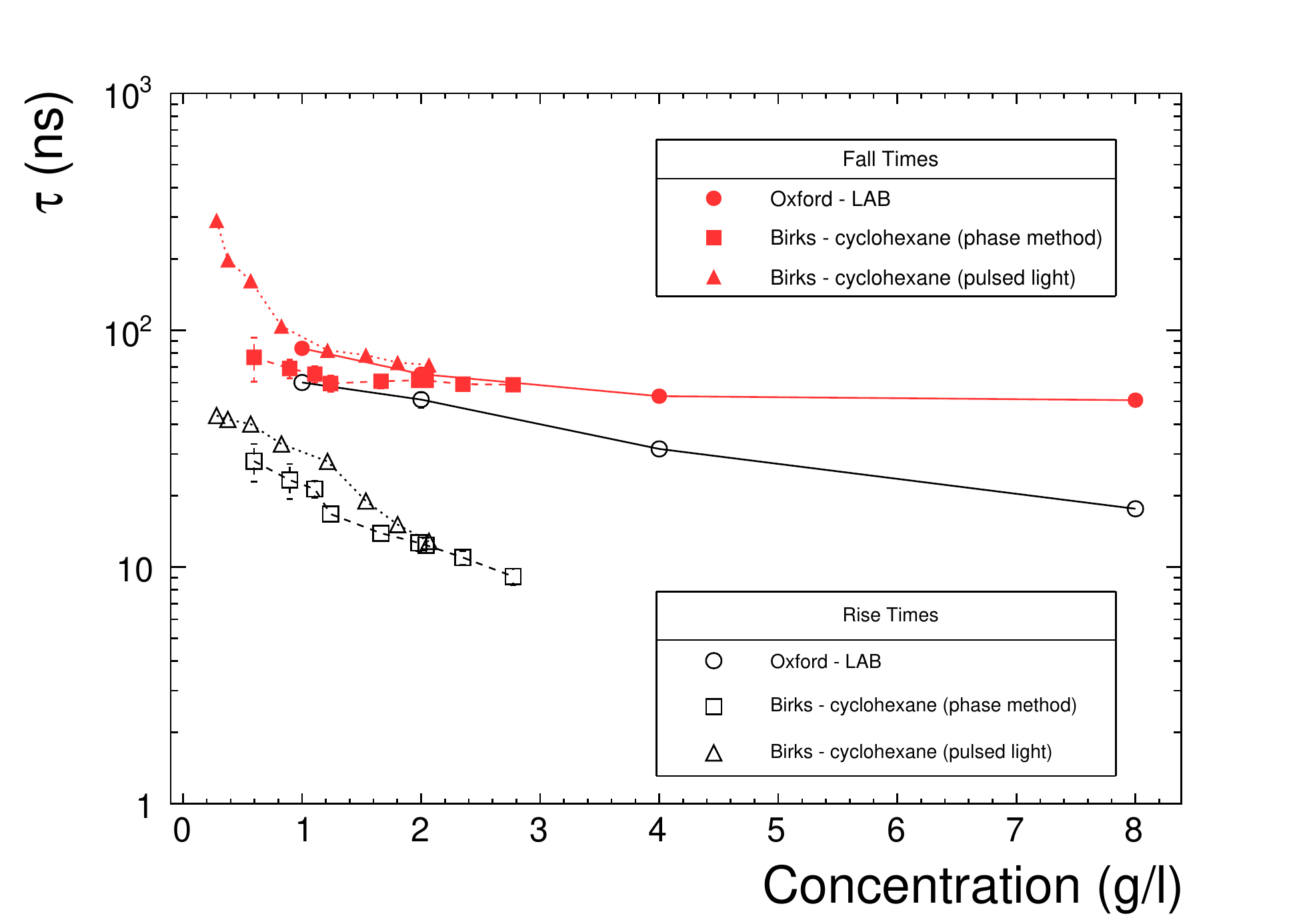}
    \caption{Excimer state}
    \label{fig:pyrene_eximer_summary}
  \end{subfigure}
  \caption{Comparison of Pyrene measurements against those given in \cite{Birks}. Rise and fall times are given in open and closed markers (black and red), respectively.}
  \label{fig:pyrene_comparison_summary}
\end{figure}

\section{9,10-Diphenylanthracene (DPA)}

\begin{figure}[H]
\centering
  \begin{subfigure}{0.7\textwidth}
    \centering
    \includegraphics[width=0.9\linewidth]{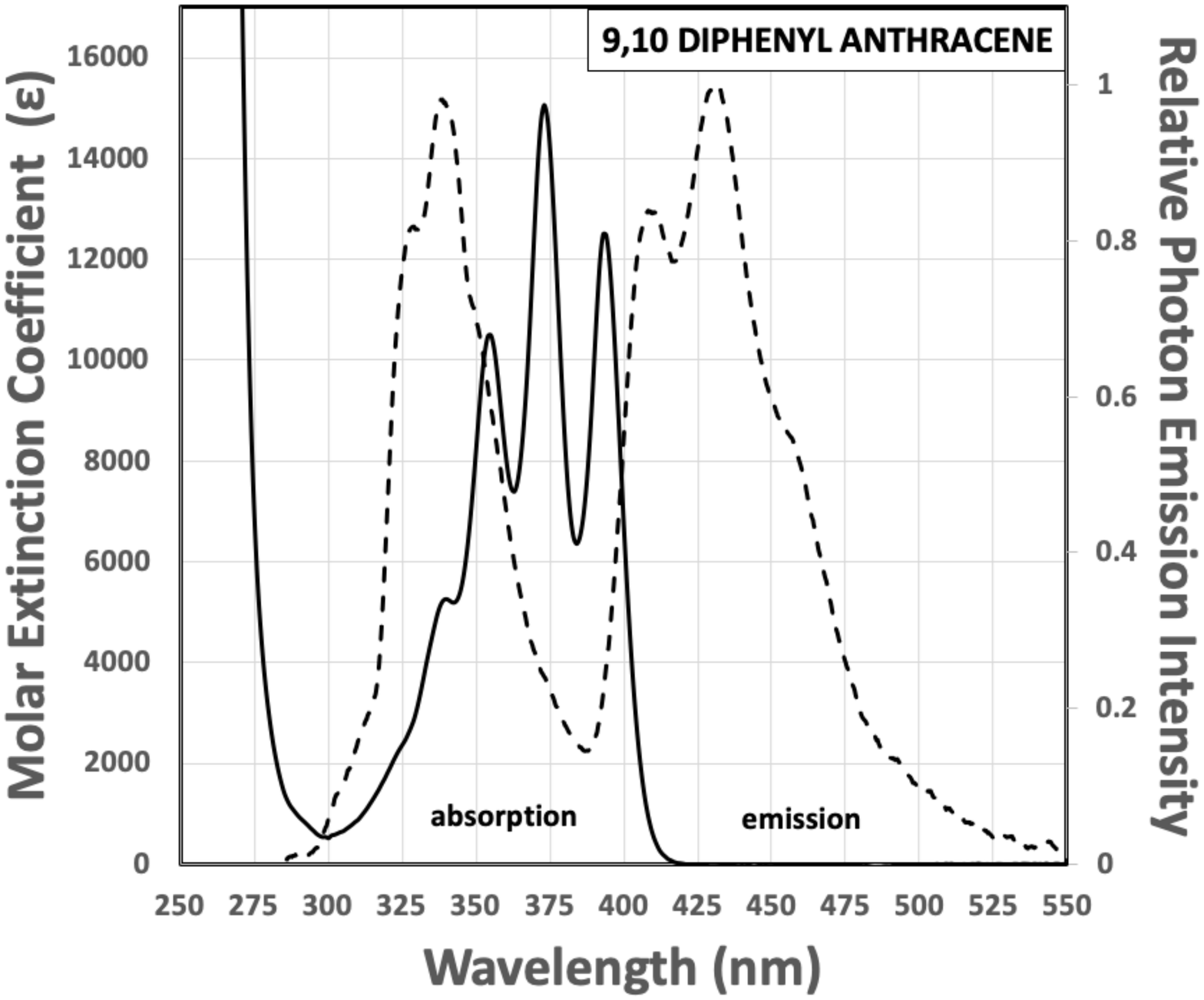}
    \caption{Absorption (in cyclohexane, solid line) and relative emission spectra (0.1g/L in LAB, dashed line)}
    \label{fig:DPA1}
  \end{subfigure} \\
  \vskip 0.3in
  \begin{subfigure}{0.7\textwidth}
    \centering
    \includegraphics[width=0.9\linewidth]{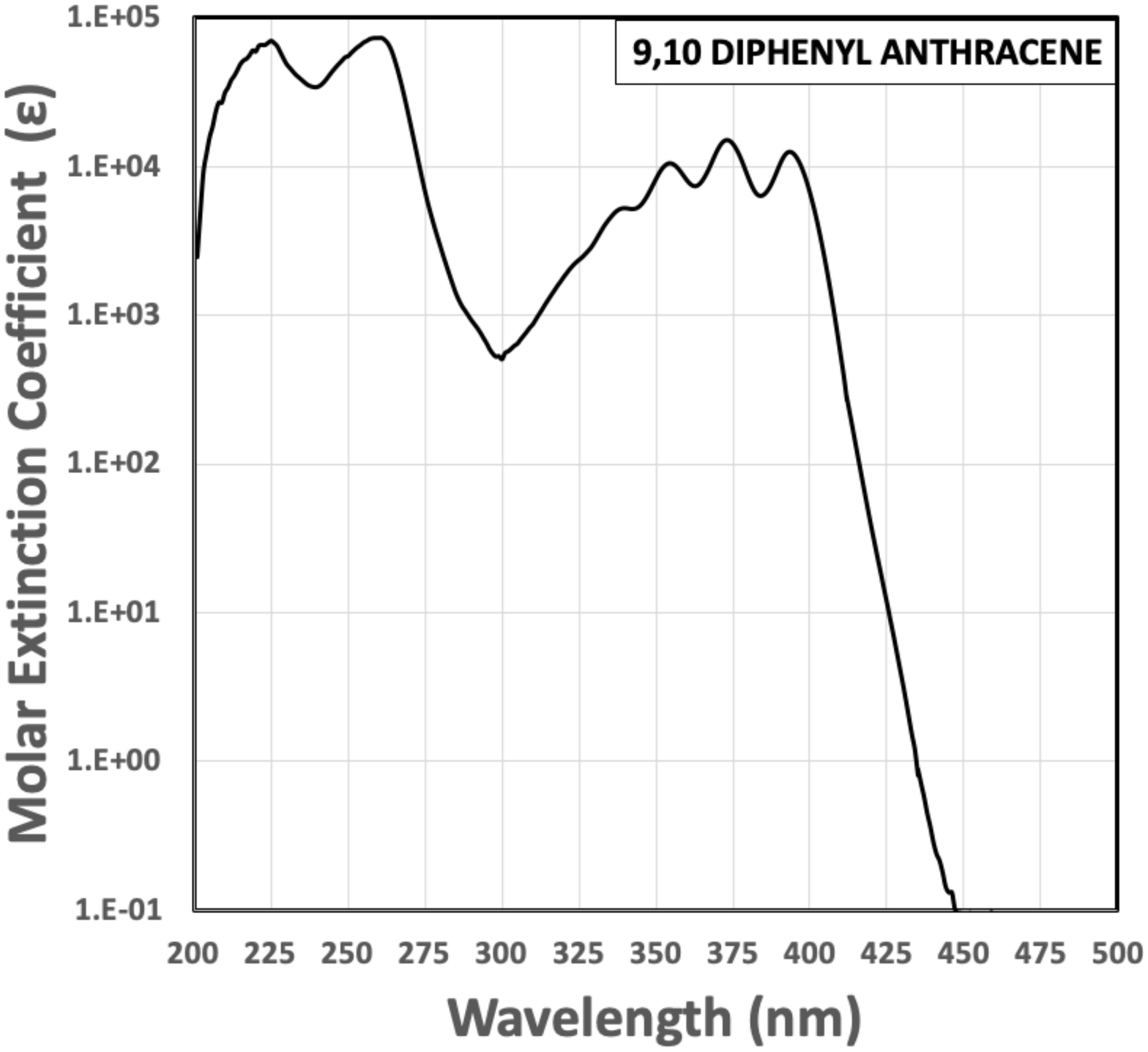}
    \caption{Logarithmic absorption spectrum}
    \label{fig:DPA2}
  \end{subfigure}
  \caption{DPA absorption spectrum in cyclohexane and relative emission spectrum in LAB.}
\end{figure}

DPA (CAS 1499-10-1) is a yellow, crystalline solid with a melting point of 250$^o$C and a chemical formula of C$_{26}$H$_{18}$ (MW 330.42~g/mol) that comprises three fused benzene rings with two additional linked rings from the centre of the chain. The DPA sample used here was obtained from Tokyo Chemical Company (TCI) with $>$98~\% purity. Figure \ref{fig:DPA1} shows the absorption and relative emission spectra in LAB, with figure \ref{fig:DPA2} showing more details of the absorption on a logarithmic scale.

When used as a secondary fluor in conjunction with 2~g/l PPO in LAB, the light yield of the mixture was found to reach $93\pm1$~\% that of the PPO reference alone, which is consistent with the high quantum yield of DPA \cite{DPA}. For measurements here, a concentration of 0.3~g/l was used so as to insure nearly complete absorption of the PPO emission spectrum within the vial while still maintaining a dominantly radiative transfer to DPA. 

It is interesting to note two distinct components of absorption and emission, with the lower wavelength absorption in the 250~nm region roughly 50 times stronger than that around 375~nm and corresponding emission near 340~nm overlapping the secondary absorption region. These features were missing the measurements by Berlman \cite{Berlman}, likely because measurements were not extended low enough in wavelength. DPA can be dissolved in LAB a concentrations as high as 5~g/l at room temperature and, in principle, could then also be used as a primary fluor, although it is several times more expensive than PPO and would suffer from notable absorption below $\sim$450~nm in large scale detectors, with a large proportion of light then shifted to less efficient detection ranges for bialkalai PMTs. As a primary fluor, we note that the decay time measured below becomes slightly longer. We believe this has to do with the lower wavelength emission peak and the subsequent transfer of energy to the overlapping absorption bands.

\begin{figure}[H]
  \begin{subfigure}{0.5\textwidth}
    \centering
    \includegraphics[width=0.9\linewidth]{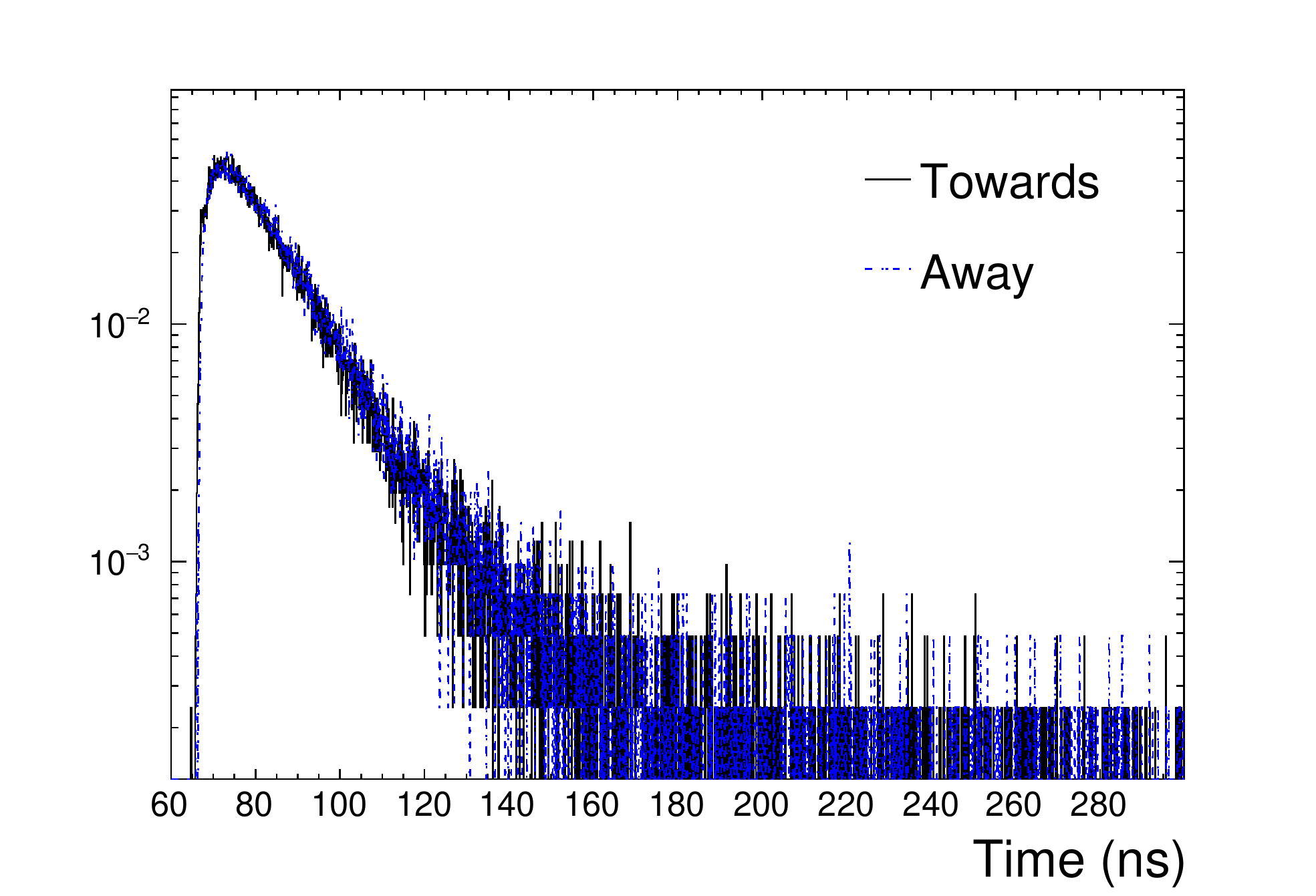}
    \caption{timing spectra (\textit{time measurement PMT})}
    \label{fig:compare_time_DPA_5gl}
  \end{subfigure}%
  \begin{subfigure}{0.5\textwidth}
    \centering
    \includegraphics[width=0.9\linewidth]{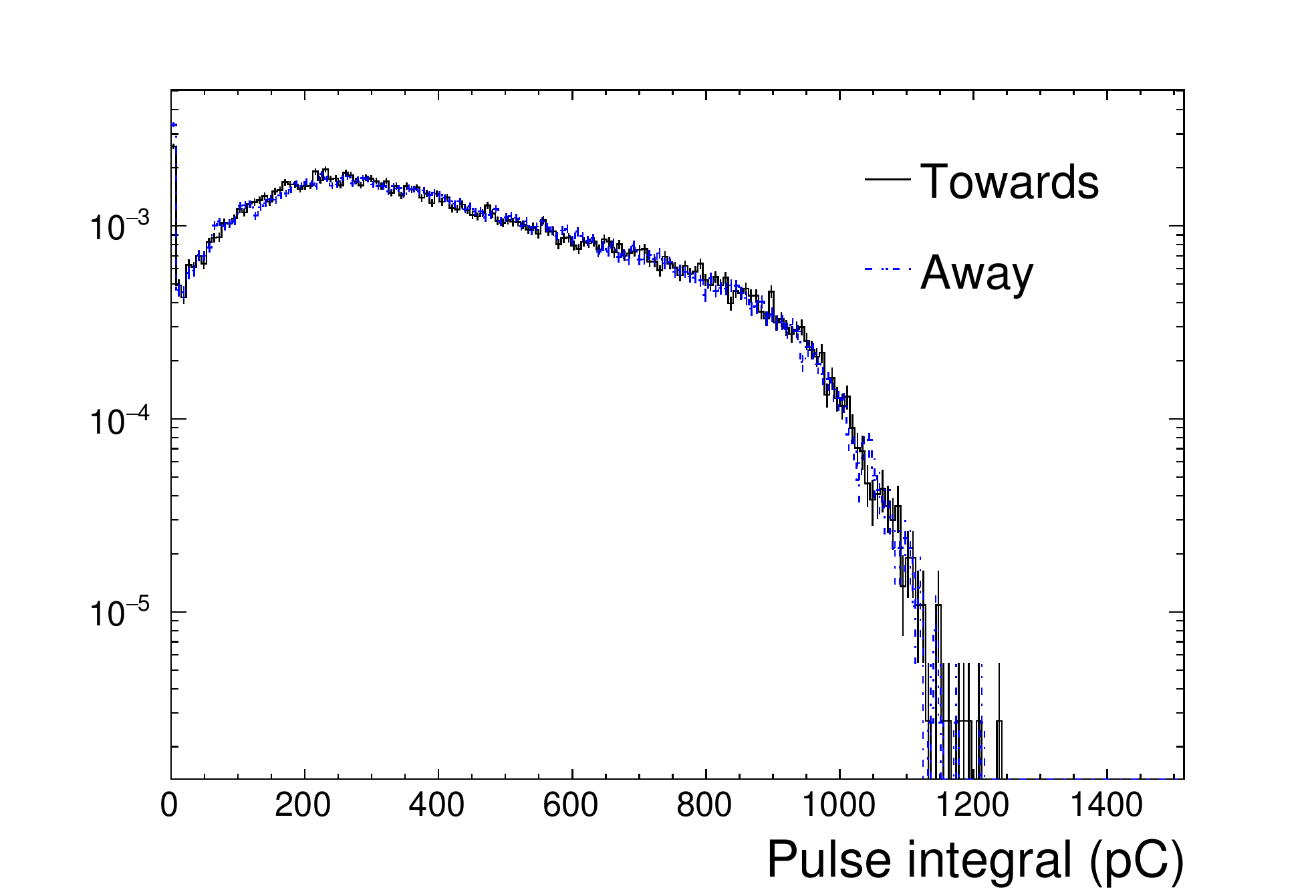}
    \caption{charge spectra (\textit{charge collection PMT})} 
    \label{fig:compare_charge_DPA_5gl}
  \end{subfigure}
  \vskip\baselineskip
  \begin{subfigure}{0.5\textwidth}
    \centering
    \includegraphics[width=0.9\linewidth]{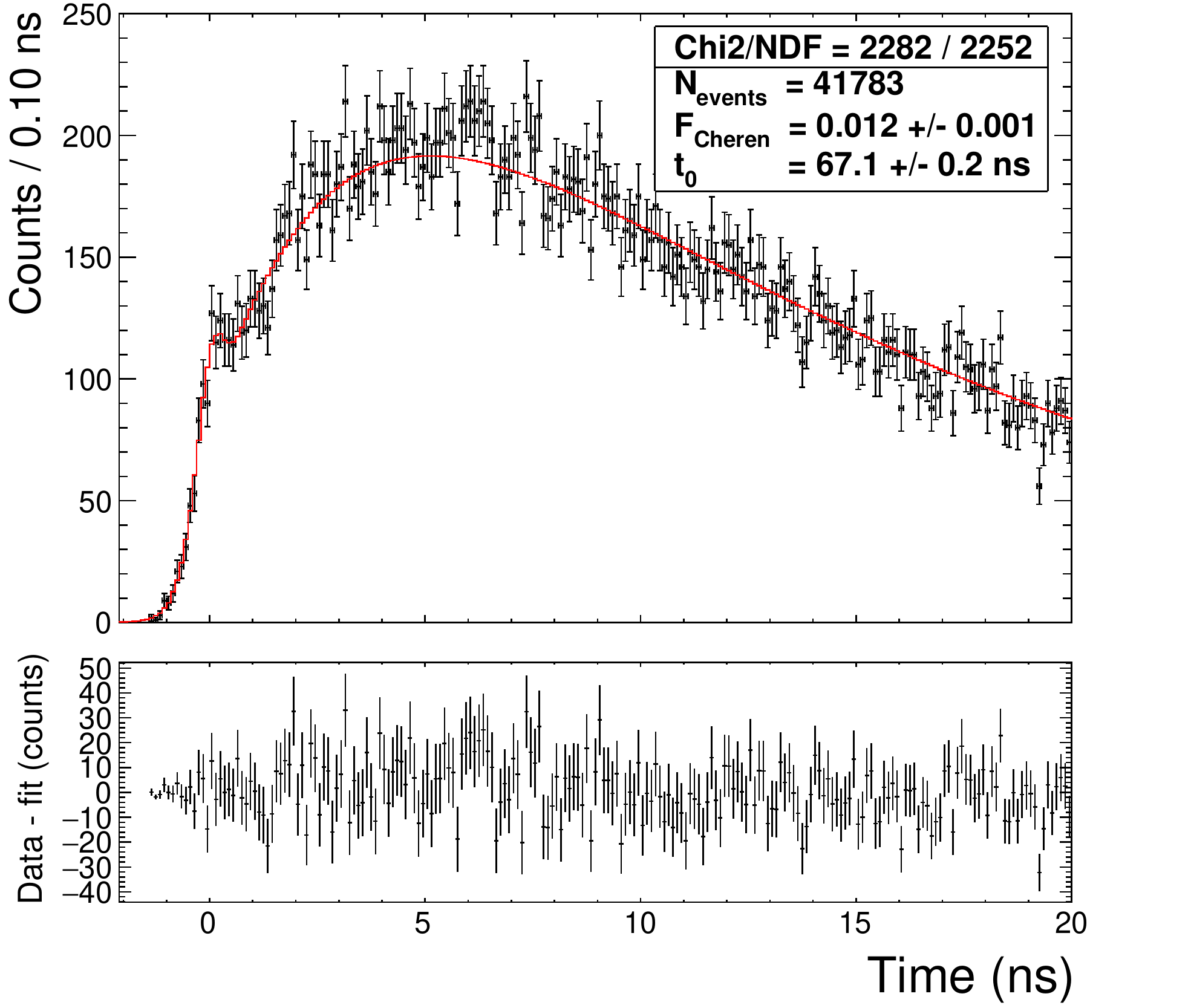}
    \caption{fit to \textit{towards} configuration}
    \label{fig:fit_DPA_5gl_towards}
  \end{subfigure}%
  \begin{subfigure}{0.5\textwidth}
    \centering
    \includegraphics[width=0.9\linewidth]{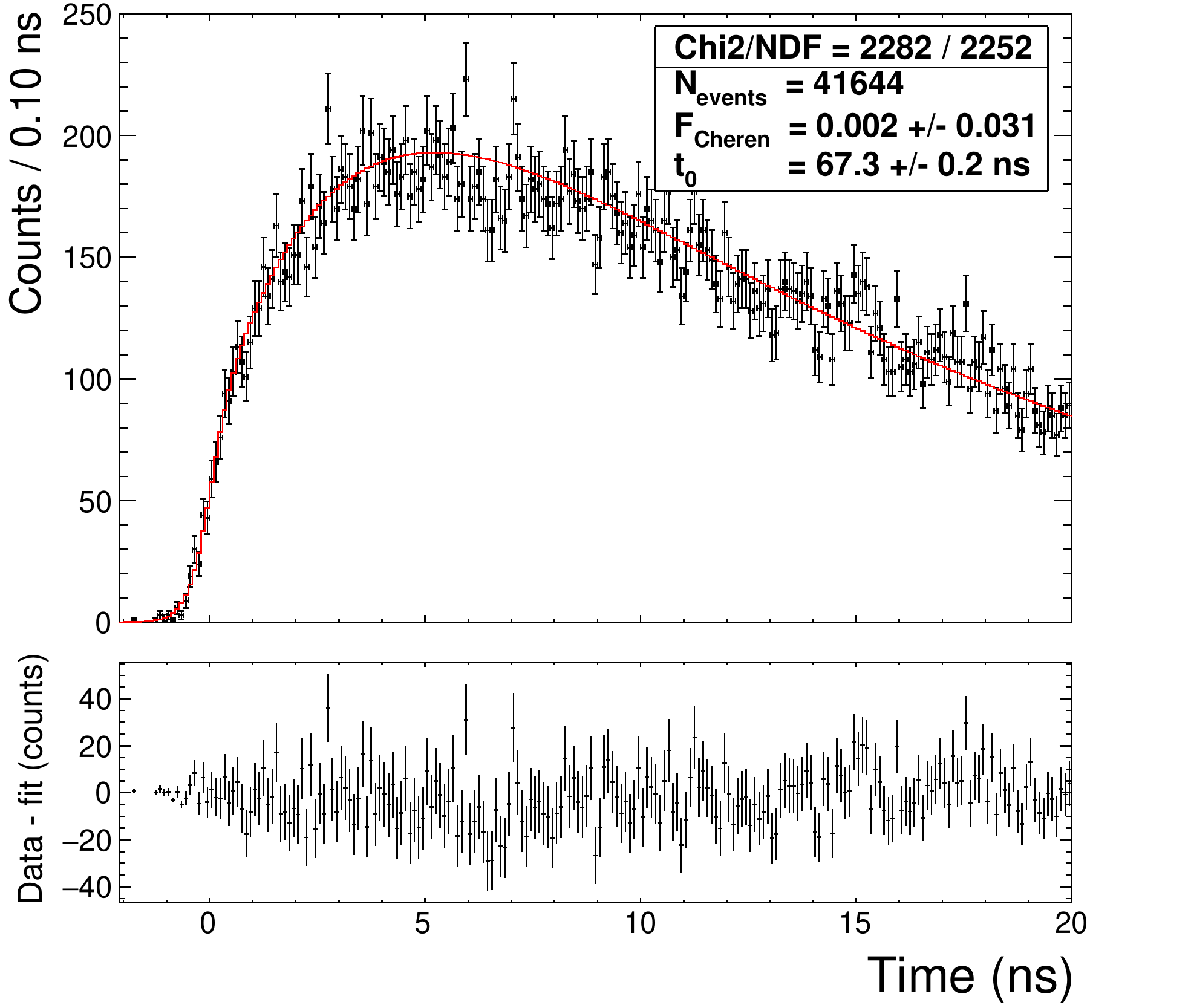}
    \caption{fit to \textit{away} configuration}
    \label{fig:fit_DPA_5gl_away}
  \end{subfigure}
  \caption{Time profile results for 5~g/l DPA. The fit parameters associated with the scintillation light are given in Tables~\ref{tab:time_constants} and \ref{tab:scale_constants}.}
\end{figure}

\begin{figure}[H]
  \begin{subfigure}{0.5\textwidth}
    \centering
    \includegraphics[width=0.9\linewidth]{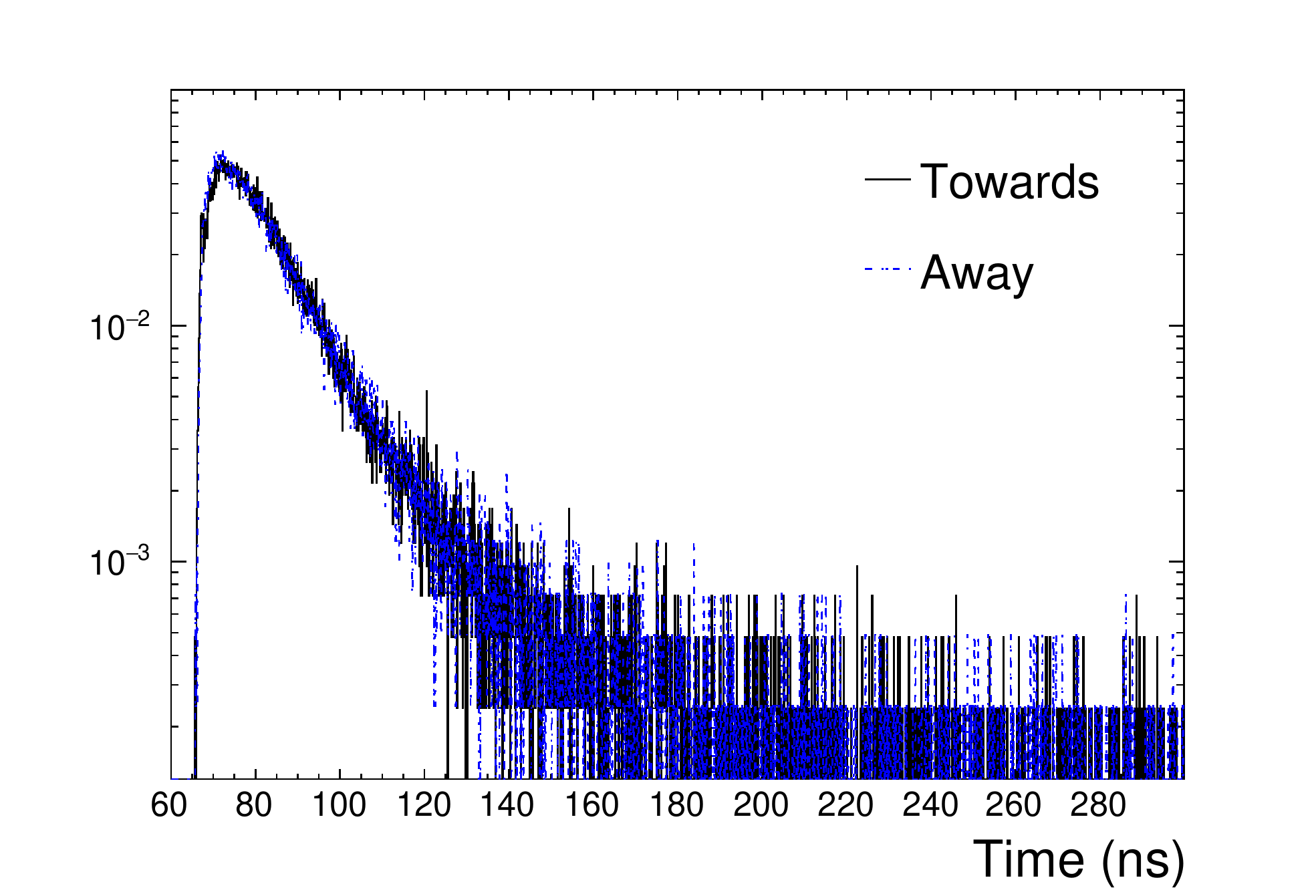}
    \caption{timing spectra (\textit{time measurement PMT})}
    \label{fig:compare_time_DPA_PPO}
  \end{subfigure}%
  \begin{subfigure}{0.5\textwidth}
    \centering
    \includegraphics[width=0.9\linewidth]{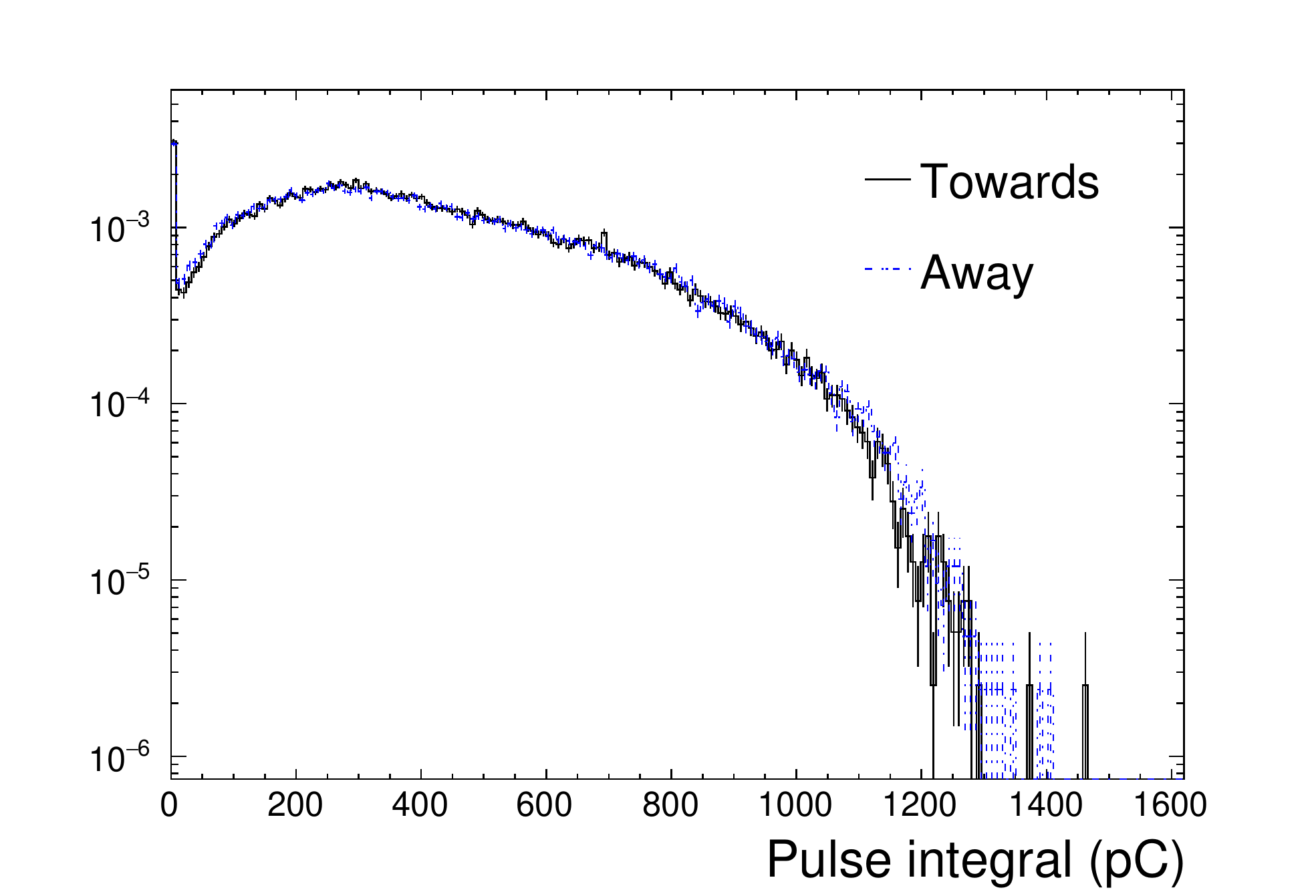}
    \caption{charge spectra (\textit{charge collection PMT})} 
    \label{fig:compare_charge_DPA_PPO}
  \end{subfigure}
  \vskip\baselineskip
  \begin{subfigure}{0.5\textwidth}
    \centering
    \includegraphics[width=0.9\linewidth]{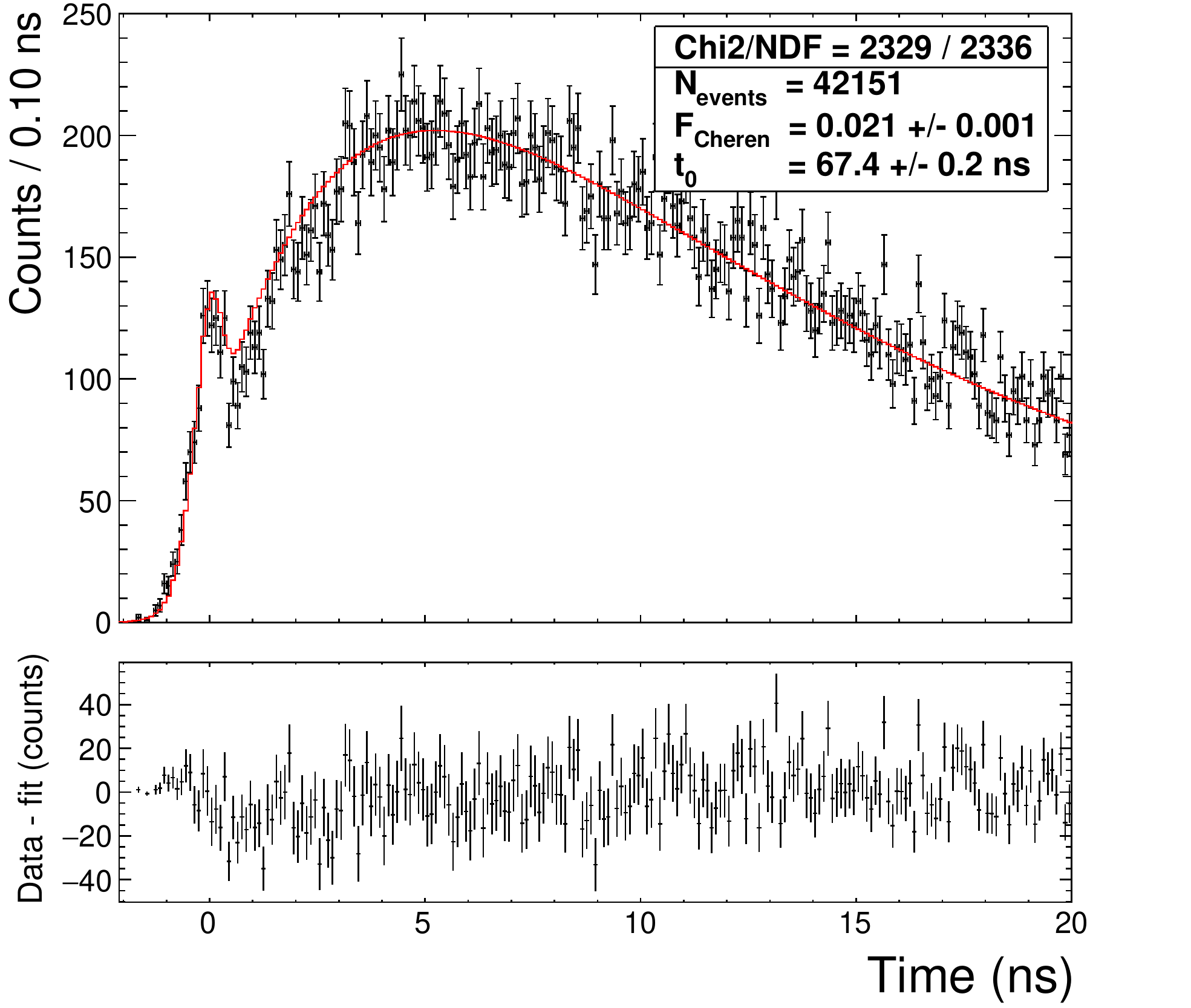}
    \caption{fit to \textit{towards} configuration}
    \label{fig:fit_DPA_PPO_towards}
  \end{subfigure}%
  \begin{subfigure}{0.5\textwidth}
    \centering
    \includegraphics[width=0.9\linewidth]{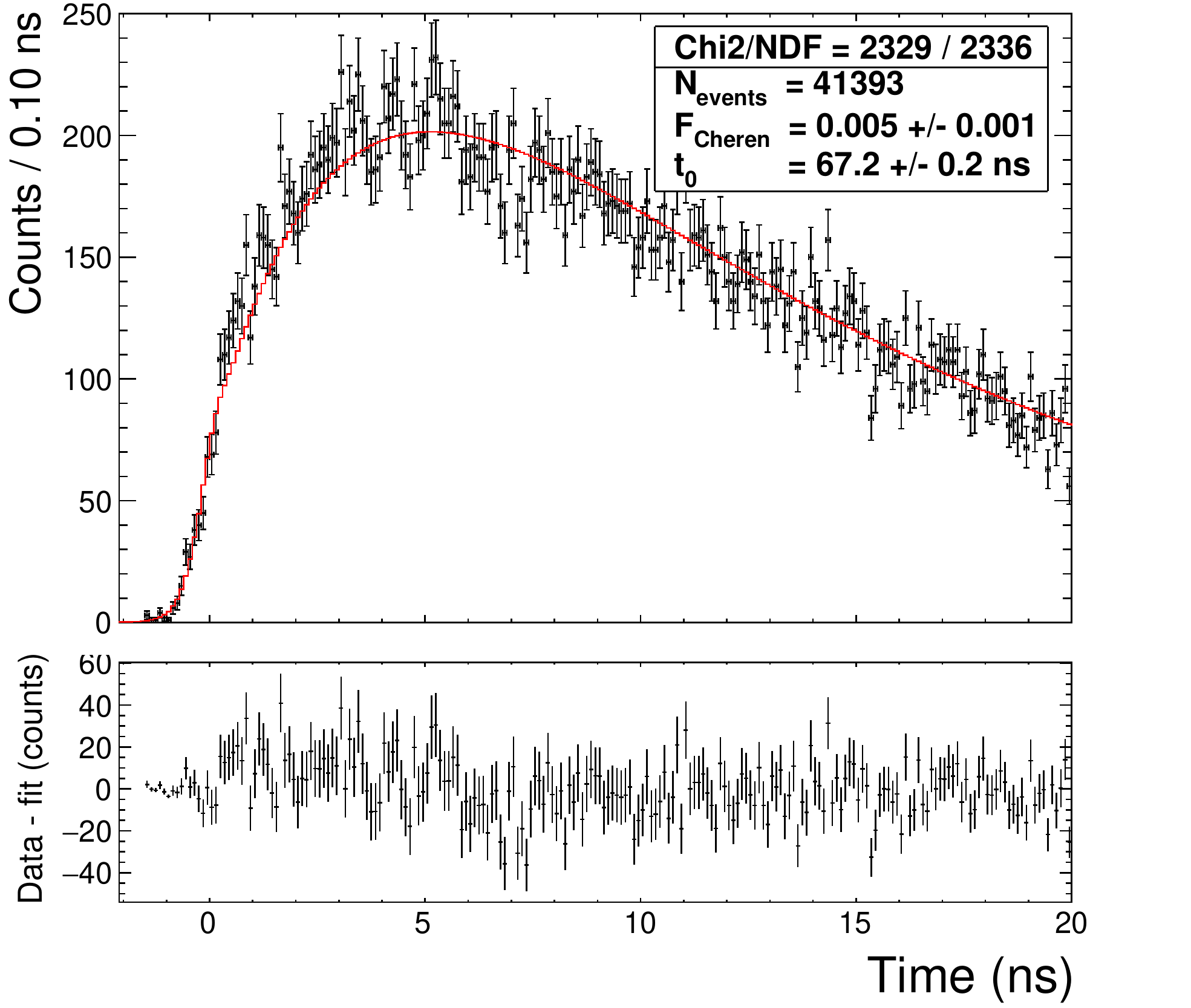}
    \caption{fit to \textit{away} configuration}
    \label{fig:fit_DPA_PPO_away}
  \end{subfigure}
  \caption{Time profile results for 0.3~g/l DPA + 2~g/l PPO. The fit parameters associated with the scintillation light are given in Tables~\ref{tab:time_constants} and \ref{tab:scale_constants}.}
\end{figure}

The timing spectra for the forward and backward experimental configurations using DPA as a primary and secondary fluor are given in Figures~13 
and 14, 
respectively. The results of fits to the measured timing spectra are given in Tables~\ref{tab:time_constants} and \ref{tab:scale_constants}, showing a rise time of $\sim$3.2 ns and a primary decay time of $\sim$12 ns. The extent of Cherenkov separation is, in fact, similar for both DPA concentrations used: the main difference between Figures~\ref{fig:fit_DPA_5gl_towards} and \ref{fig:fit_DPA_PPO_towards} is that the scintillation signal in the former is $\sim$20~\% higher owing to the better transfer efficiency of the higher concentration fluor and its greater quantum efficiency compared to the PPO-DPA combination. The Cherenkov separation is still good, but the contamination from scintillation light is greater than for acenapthene or pyrene owing to the faster decay time. This contamination will increase in large detectors owing to dispersion effects. However, this also allows for improved vertex reconstruction and less susceptibility to fluorescence quenching in loaded scintillator mixtures (which tends to increase with fluor lifetime). Any fluorescence quenching that is present will tend to improve the visibility of the Cherenkov signal again, so this may be the better choice of fluor for certain physics applications.


\section{1,6-Diphenyl-1,3,5-hexatriene (DPH)}

\begin{figure}[H]
\centering
  \begin{subfigure}{0.8\textwidth}
    \centering
    \includegraphics[width=0.9\linewidth]{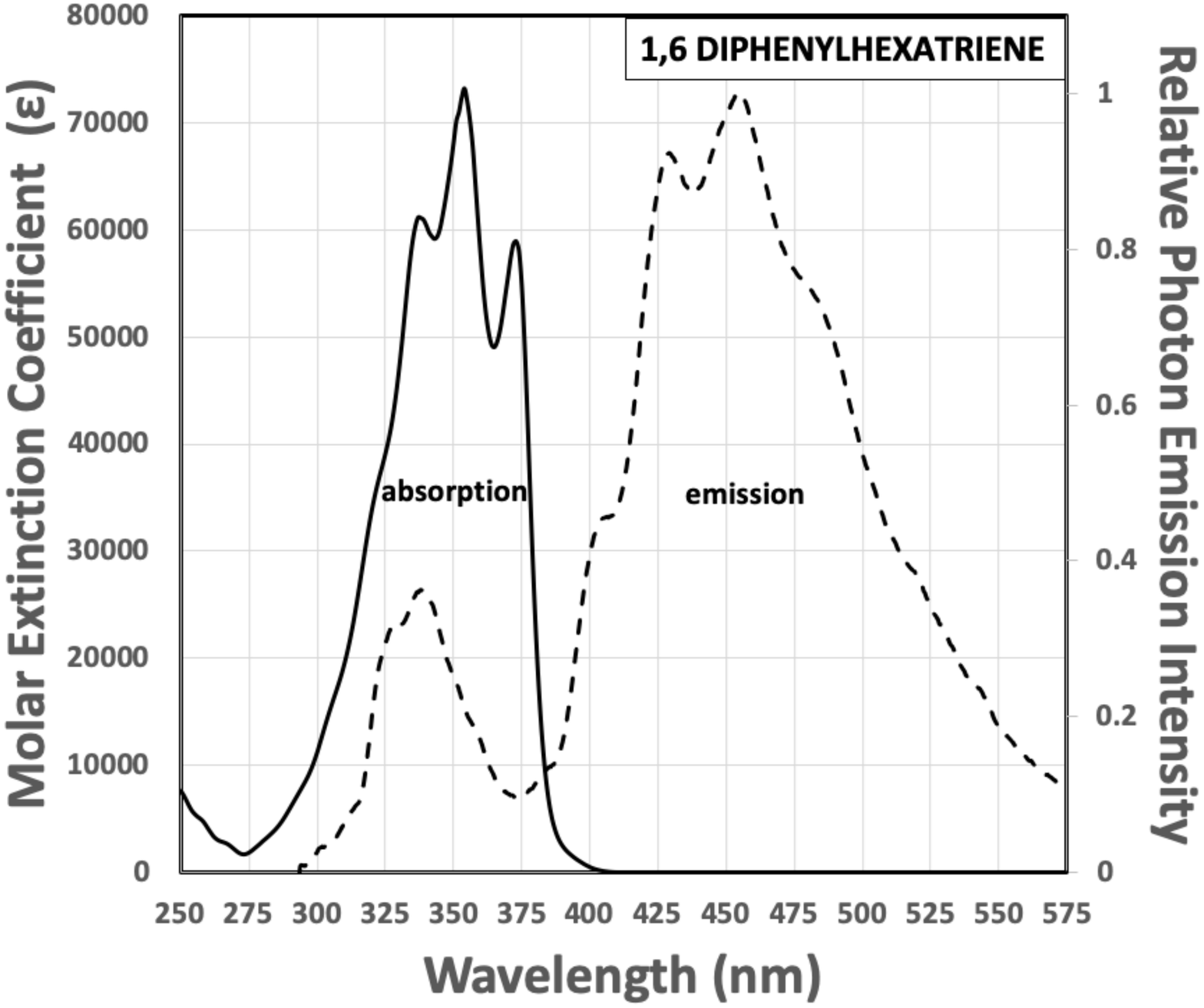}
    \caption{Absorption (in cyclohexane, solid line) and relative emission spectra (0.1 g/L in LAB, dashed line)}
    \label{fig:DPH1}
  \end{subfigure}\\
  \vskip 0.3in
  \begin{subfigure}{0.8\textwidth}
    \centering
    \includegraphics[width=0.9\linewidth]{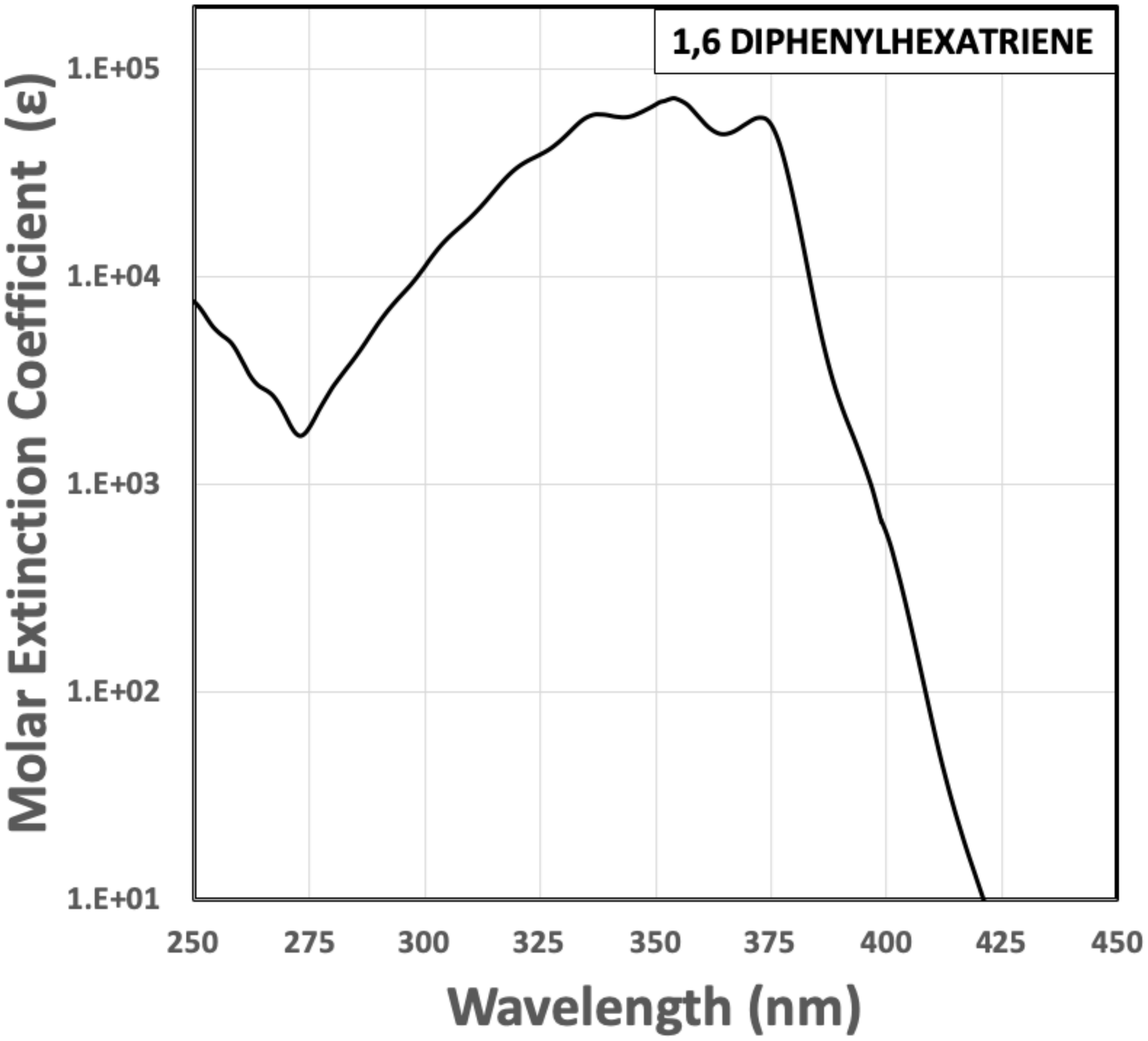}
    \caption{Logarithmic absorption spectrum}
    \label{fig:DPH2}
  \end{subfigure}
  \caption{DPH absorption spectrum in cyclohexane and relative emission spectrum in LAB.}
\end{figure}

DPH (CAS 17329-15-6) is a yellow, crystalline solid with a melting point of 200$^o$C and a chemical formula of C$_{18}$H$_{16}$ (MW 232.326~g/mol) that comprises two benzene rings connected by a hexatriene chain. The DPH sample used here was obtained from Tokyo Chemical Company (TCI) with $>$95~\% purity. Figure \ref{fig:DPH1} shows the absorption and relative emission spectra in LAB, with figure \ref{fig:DPH2} showing more details of the absorption on a logarithmic scale.

When used as a secondary fluor in conjunction with 2~g/l PPO in LAB, the light yield of the mixture was found to reach $99\pm6$~\% that of the PPO reference alone, which is a little high but consistent within the errors of that expected for the quantum yield range typically quoted for DPH \cite{DPH} (it should also be noted that literature values for quantum yields tend to vary by roughly $\sim$10~\% for the fluors considered here). A concentration of 0.1~g/l was used so as to insure nearly complete absorption of the PPO emission spectrum within the vial while still maintaining radiative transfer to DPH. As a significant portion of the emission spectrum lies above 450~nm, where bialkalai photocathodes begin to become less efficient, observed light levels in typical large format PMTs will tend to be $\sim$25~\% lower relative to PPO. DPH is roughly 5 times more expensive than DPA, but much less is needed as a secondary fluor owing to the significantly higher molar absorption coefficient.

The timing spectra for the forward and backward experimental configurations using DPH as a secondary fluor are given in Figure~16. 
The results of fits to the measured timing spectra are given in Tables~\ref{tab:time_constants} and \ref{tab:scale_constants}, showing a rise time of $2.2\pm0.2$ ns and a primary decay time of $11.4\pm0.4$ ns. The Cherenkov separation is very similar to that for DPA.

\begin{figure}[H]
  \begin{subfigure}{0.5\textwidth}
    \centering
    \includegraphics[width=0.9\linewidth]{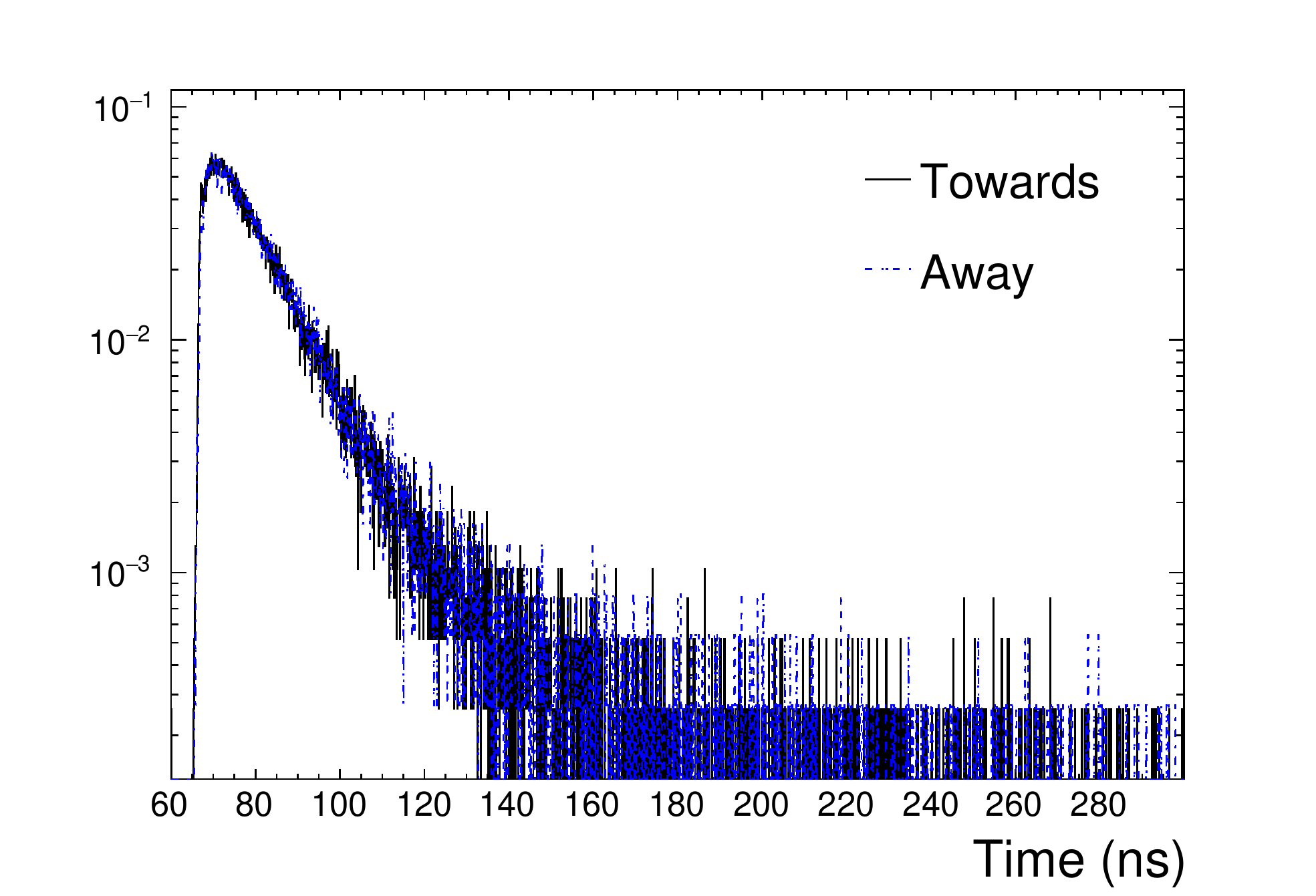}
    \caption{timing spectra (\textit{time measurement PMT})}
    \label{fig:compare_time_DPH}
  \end{subfigure}%
  \begin{subfigure}{0.5\textwidth}
    \centering
    \includegraphics[width=0.9\linewidth]{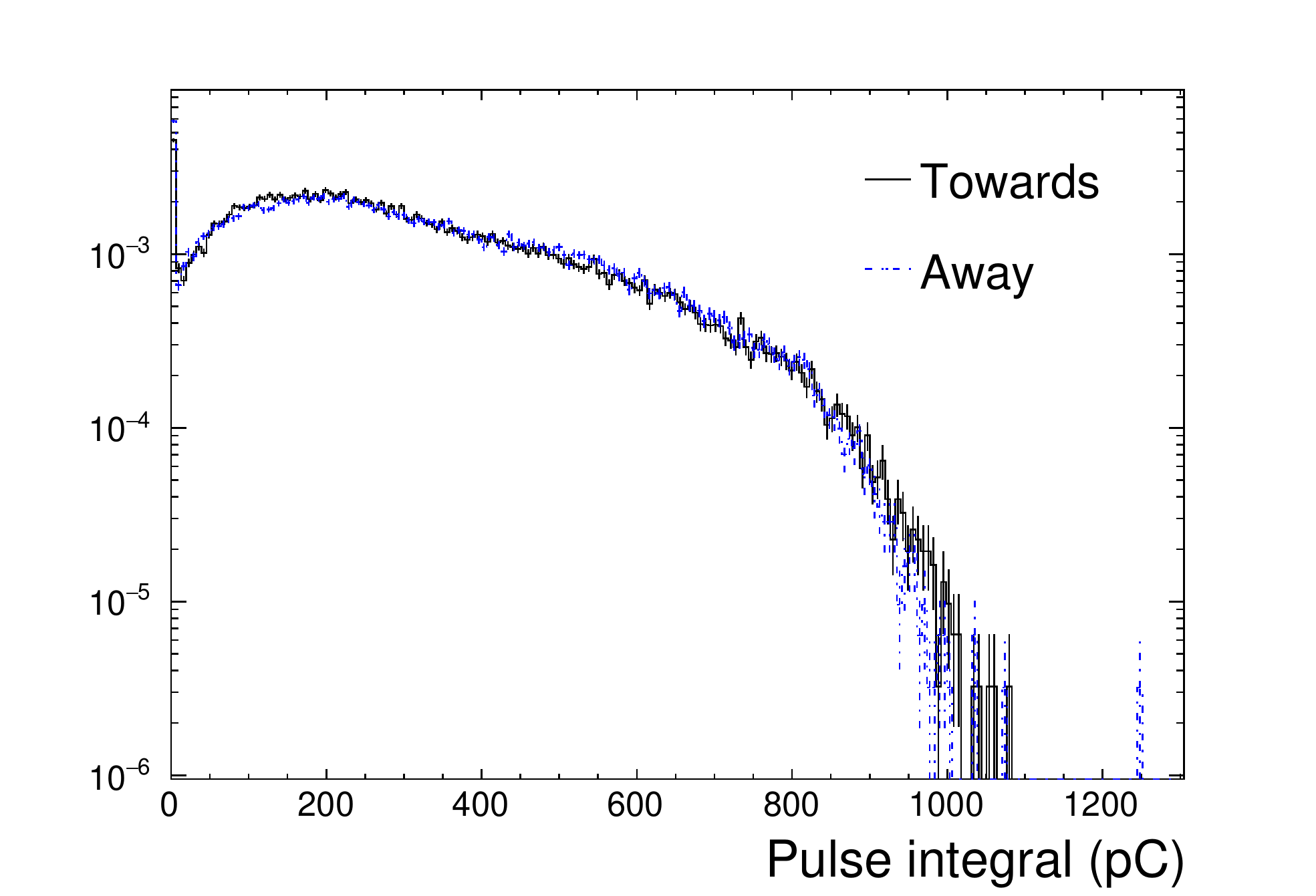}
    \caption{charge spectra (\textit{charge collection PMT})} 
    \label{fig:compare_charge_DPH}
  \end{subfigure}
  \vskip\baselineskip
  \begin{subfigure}{0.5\textwidth}
    \centering
    \includegraphics[width=0.9\linewidth]{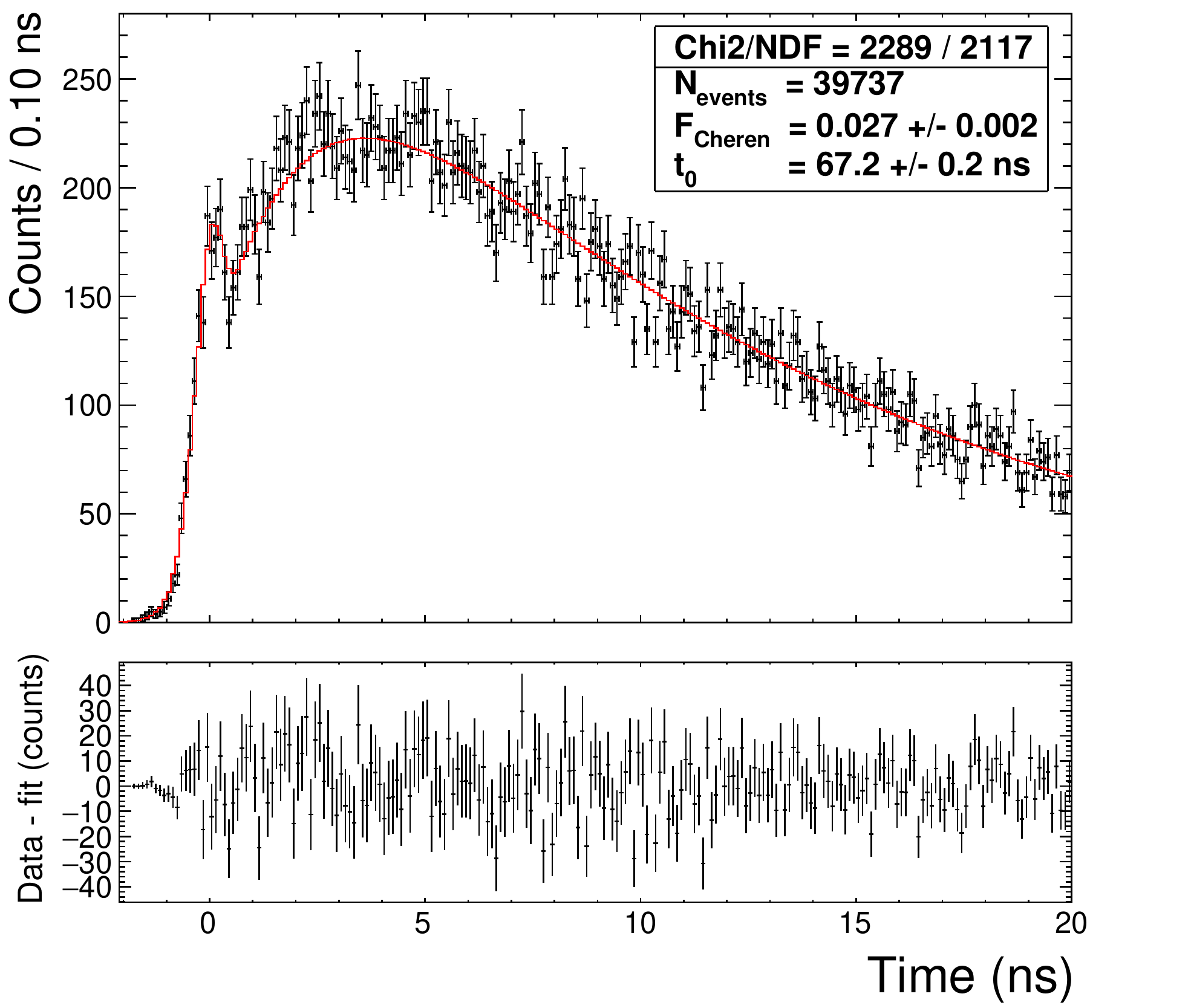}
    \caption{fit to \textit{towards} configuration}
    \label{fig:fit_DPH_towards}
  \end{subfigure}%
  \begin{subfigure}{0.5\textwidth}
    \centering
    \includegraphics[width=0.9\linewidth]{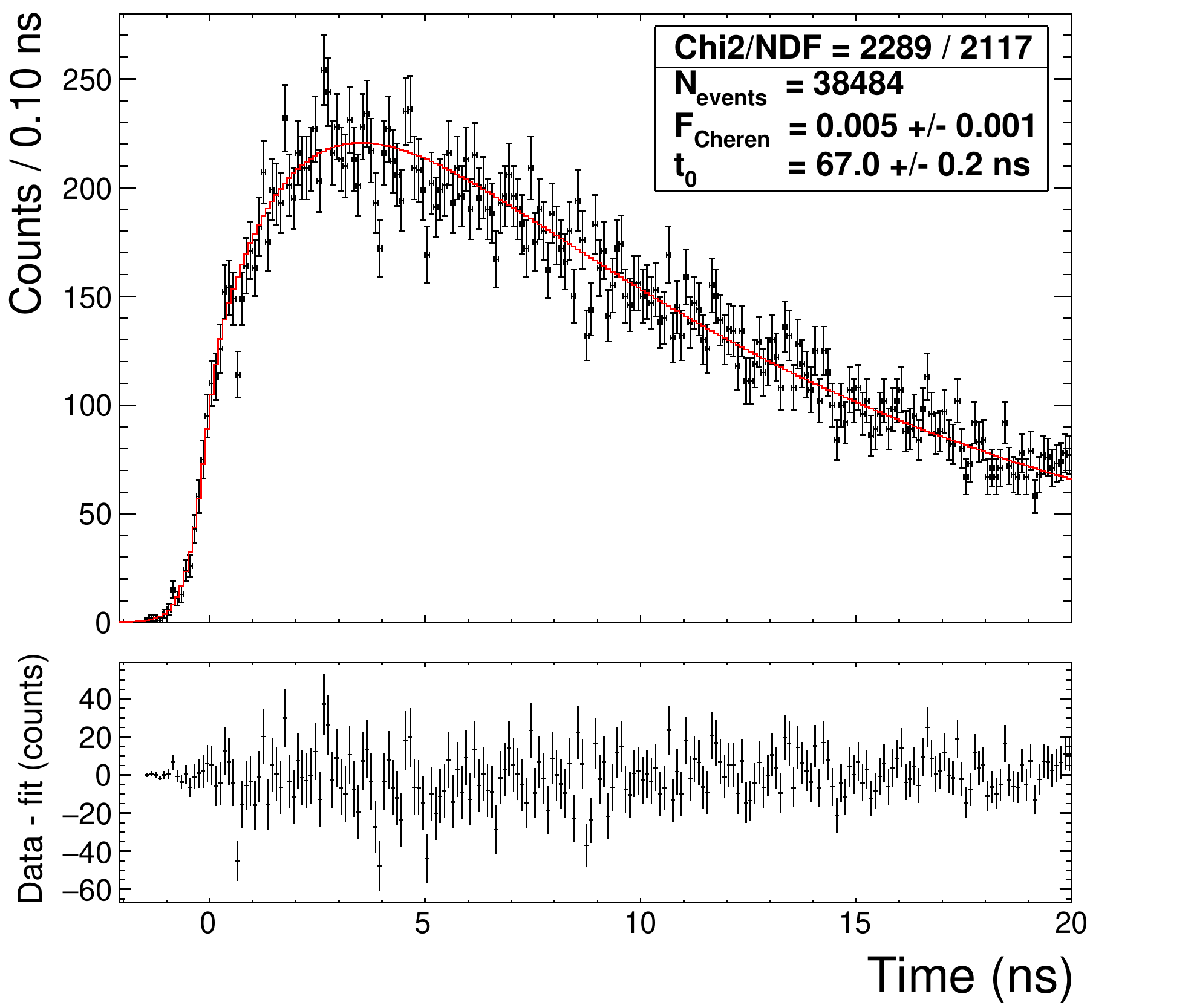}
    \caption{fit to \textit{away} configuration}
    \label{fig:fit_DPH_away}
  \end{subfigure}
  \caption{Time profile results for 2~g/l DPH. The fit parameters associated with the scintillation light are given in Tables~\ref{tab:time_constants} and \ref{tab:scale_constants}.}
\end{figure}

\section{Summary Tables}
\begin{table}[H]
\centering
\begin{tabular}{|l|c|c|c|c|c|c|}
\hline
Fluor & Conc. (g/l) & Peak em (nm) & QY & Intrinsic LY in LAB  \\ 
 &   & & & (photons/MeV) \\ \hline \hline
PPO & 2 & 360 & 0.8 \cite{PPO},\cite{PPO2} & 11900 \cite{Penn}  \\ \hline
Acenaphthene & 4 & 335 & 0.6 \cite{Berlman}& 7686$~\pm~$315  \\ \hline
Pyrene & 1 & 390(m),480(e) & 0.65 \cite{pyrene} & 9430$~\pm~$509 \\ 
 & 10 & 480(e) & & 11833$~\pm~$766   \\ \hline
DPA & 5 & 430 & 0.95 \cite{DPA} & 13584$~\pm~$582 \\ \hline
DPA+PPO & 0.3, 2 & 430 & 0.95 \cite{DPA} & 11610$~\pm~$498 \\ \hline
DPH+PPO & 0.1, 2 & 450 & 0.71 \cite{DPH} & 12356$~\pm~$926 \\ \hline
\end{tabular}
\caption{Peak emission wavelengths and intrinsic light yields for various fluors in LAB. Uncertainties are dominated by systematics due to varying vial glass thickness and sample preparation.}
\label{tab:fluors}
\end{table}

\begin{table}[H]
\centering
\begin{adjustbox}{width=1\textwidth}
\small
\begin{tabular}{|c|c|c|c|c|c|c|} \hline
Fluor & Conc. (g/l) & $\tau_{rise}$ (ns) & $\tau_{1}$ (ns) & $\tau_{2}$ (ns) & $\tau'$ (ns) & $\chi^2 / dof$ \\ 
 \hline
 \hline
DPA & 5.0 & $3.2~\pm~0.3$ & $13.0~\pm~0.2$ & $76.3~\pm~18.0$ & $3.2~\pm~1.3$ & 2311 / 2243 \\
DPA + PPO & 0.3, 2.0 &  $3.4~\pm~0.3$ &  $11.2~\pm~0.3$ & $49.0~\pm~5.6$ & $2.9~\pm~1.4$  & 2319 / 2334 \\
DPH + PPO & 0.1, 2.0 & $2.2~\pm~0.2$ & $11.4~\pm~0.4$ & $67.1~\pm~21.0$ & $3.4~\pm~1.2$ & 2250 / 2107 \\
Acenapthene & 4.0 & $2.1~\pm~0.2$ & $45.4~\pm~0.3$ & - & $0.9~\pm~0.4$ & 4161 / 3987 \\
\hline
\multirow{5}{*}{Pyrene (excimer)} 
& 1.0 & $60.1~\pm~1.2$ & $83.8~\pm~1.3$ & - & $0.5~\pm~2.6$ & 7962 / 8088 \\
& 2.0 & $50.9~\pm~3.8$ & $65.2~\pm~3.6$ & - & $0.6~\pm~0.2$ & 7861 / 7375 \\
& 4.0 & $31.5~\pm~1.0$ & $52.6~\pm~1.0$ & - & $0.5~\pm~0.1$ & 6114 / 5739 \\
& 8.0 & $17.6~\pm~0.5$ & $50.6~\pm~0.6$ & - & $2.4~\pm~13.5$  & 5968 / 4968 \\
\hline
\multirow{2}{*}{Pyrene (monomer)}
& 1.0 & $4.6~\pm~1.7$ & $101.2~\pm~0.6$ & - & $2.1~\pm~7.6$ & 7084 / 7059 \\
& 2.0 & $4.5~\pm~0.8$ & $63.8~\pm~0.5$ & - & $7.0~\pm~0.8$ & 5450 / 5378 \\
\hline
\end{tabular}
\end{adjustbox}
\caption{Best fit rise time and decay constants using the procedure described in section 3. }
\label{tab:time_constants}
\end{table}

\begin{table}[H]
\centering
\begin{adjustbox}{width=1\textwidth}
\small
\begin{tabular}{|c|c|c|c|c|} \hline
Fluor & Concentration (g/l) & $A_{1}$ (\%) & $A_{2}$ (\%) & $A'$ (\%) \\ 
 \hline
 \hline
DPA & 5.0 & $89.7~\pm~2.5$ & $5.7~\pm~3.3$ & $4.5~\pm~2.1$ \\
DPA + PPO & 0.3, 2.0  & $85.3~\pm~2.9$ & $10.3~\pm~3.6$ & $4.5~\pm~2.2$ \\
DPH + PPO & 0.1, 2.0  & $85.6~\pm~2.3$ & $5.8~\pm~3.5$ & $8.6~\pm~2.6$ \\
Acenapthene & 4.0 & $98.6~\pm~0.3$ & - &  $1.4~\pm~0.3$\\
\hline
\multirow{5}{*}{Pyrene (excimer)} 
& 1.0 & $99.2~\pm~0.02$ & - & $0.8~\pm~0.2$ \\
& 2.0 & $99.8~\pm~0.1$ & - & $0.2~\pm~0.1$ \\
& 4.0 & $99.7~\pm~0.1$ & - & $0.3~\pm~0.1$ \\
& 8.0 & $100.0~\pm~0.1$ & - & $0.0~\pm~0.1$ \\
\hline
\multirow{2}{*}{Pyrene (monomer)}
& 1.0 & $99.0~\pm~1.0$ & - & $1.0~\pm~1.0$  \\
& 2.0 & $96.6~\pm~1.9$ & - & $3.4\pm~1.9$  \\
\hline
\end{tabular}
\end{adjustbox}
\caption{Best fit normalisations for scintillation components, as defined by Equation~\ref{eq:optics} and Table~\ref{tab:fit_parameters}, expressed as percentages. }
\label{tab:scale_constants}
\end{table}

\section{Conclusions}
The properties of four slow fluors have been studied in the context of LAB-based liquid scintillator mixtures to provide a means to effectively separate Cherenkov light in time from the scintillation signal with high efficiency. This allows for directional and particle ID information while also maintaining good energy resolution. Such an approach is highly economical and can be readily applied to existing and planned large-scale liquid scintillator instruments. Using this technique, this paper explicitly demonstrates Cherenkov separation on a bench-top scale, showing clear directionality, for electron energies extending below 1 MeV. This has important consequences for a variety of future instruments, including measurements of low energy solar neutrinos and searches for neutrinoless double beta decay in loaded scintillator detectors. This also opens the possibility of obtaining good directional information for elastic scattering events from supernovae neutrinos and reactor anti-neutrinos in large scale liquid scintillation detectors. While the use of slow fluors means that the vertex resolution may be worse than typical large scale liquid scintillator detectors (but better than typical large scale Cherenkov detectors), the balance between position resolution, Cherenkov separation purity and energy resolution can be tuned for a particular physics objective by modifying the fluor mixture. This balance is also affected by the presence of fluorescence quenchers, which may be naturally present in the case of loaded scintillator mixtures or could be purposely introduced.

\section{Acknowledgements}
We wish to acknowledge Robert Taylor's group at Oxford university, in particular Tim Puchtler, Claudius Kochler and Mo Li, for help with the spectral measurements, early explorations of some fluorescence time measurements, and several useful discussions. The work of the authors has been supported by the Science Technology and Facilities Council (STFC) of the United Kingdom, grant number ST/S000933/1. Equipment used for the spectral and transmission measurements was also supported by the Engineering and Physical Sciences Research Council (ERPSRC), grant number EP/M012379/1.

\end{document}